\documentclass[showpacs,prx,onecolumn,superscriptaddress, nofootinbib]{revtex4-2}

\usepackage[utf8]{inputenc} 
\usepackage{qcircuit}
\usepackage[dvips]{graphicx}
\usepackage{siunitx}
\usepackage{amsmath,amssymb,amsthm,mathrsfs,amsfonts,dsfont}
\usepackage{subfigure, epsfig}
\usepackage{braket}
\usepackage{bm}
\usepackage{enumerate}
\usepackage[dvipsnames]{xcolor}
\usepackage{color}
\usepackage{fancybox, graphicx}
\usepackage[skins]{tcolorbox}
\usepackage{multirow}
\usepackage{mathtools}

\usepackage[colorlinks]{hyperref}
\hypersetup{
	colorlinks=true,
	citecolor = {blue},
    linkcolor=blue,
    filecolor=magenta,      
    urlcolor=cyan,
}
\usepackage{cleveref}

\newcommand{\eq}[1]{Eq.\ (\ref{eq:#1})}
\newcommand{\Eq}[1]{Eq.\ (\ref{Eq:#1})}
\newcommand{\thm}[1]{\hyperref[thm:#1]{Theorem~\ref*{thm:#1}}}
\newcommand{\defn}[1]{\hyperref[defn:#1]{Definition~\ref*{defn:#1}}}
\newcommand{\lem}[1]{\hyperref[lem:#1]{Lemma~\ref*{lem:#1}}}
\newcommand{\prop}[1]{\hyperref[prop:#1]{Proposition~\ref*{prop:#1}}}
\newcommand{\fig}[1]{\hyperref[Fig:#1]{Figure~\ref*{Fig:#1}}}
\newcommand{\tab}[1]{\hyperref[tab:#1]{Table~\ref*{tab:#1}}}
\renewcommand{\sec}[1]{\hyperref[Sec:#1]{Section~\ref*{Sec:#1}}}
\newcommand{\append}[1]{\hyperref[App:#1]{Appendix~\ref*{App:#1}}}
\newcommand{\cor}[1]{\hyperref[cor:#1]{Corollary~\ref*{cor:#1}}}
\newcommand{\obs}[1]{\hyperref[obs:#1]{Observation~\ref*{obs:#1}}}

\newcommand{\sfs}[1]{\big| #1 \big|_{\eta}}
\newcommand{\defindex}[1]{_{#1}}

\definecolor{amethyst}{rgb}{0.6, 0.4, 0.8}

\newcommand{\norm}[1]{\left\lVert#1\right\rVert}
\newcommand{\vertiii}[1]{{\left\vert\kern-0.25ex\left\vert\kern-0.25ex\left\vert #1
		\right\vert\kern-0.25ex\right\vert\kern-0.25ex\right\vert}}

\usepackage{mathrsfs}

\allowdisplaybreaks

\usepackage{titlesec}
\begin{document}

\title{Exploiting fermion number in factorized decompositions of the electronic structure Hamiltonian}

\author{Sam McArdle}
\affiliation{AWS Center for Quantum Computing, Pasadena, CA 91125, USA}

\author{Earl Campbell}
\affiliation{AWS Center for Quantum Computing, Cambridge, UK}

\author{Yuan Su}
\affiliation{Institute for Quantum Information and Matter, Caltech, Pasadena, CA 91125, USA}
\affiliation{Google Research, Venice, CA 90291, USA}


\begin{abstract}
Achieving an accurate description of fermionic systems typically requires considerably many more orbitals than fermions. Previous resource analyses of quantum chemistry simulation often failed to exploit this low fermionic number information in the implementation of Trotter-based approaches and overestimated the quantum-computer runtime as a result. They also depended on numerical procedures that are computationally too expensive to scale up to large systems of practical interest. Here we propose techniques that solve both problems by using various factorized decompositions of the electronic structure Hamiltonian. We showcase our techniques for the uniform electron gas, finding substantial (over $100\times$) improvements in Trotter error for low-filling fraction and pushing to much higher numbers of orbitals than is possible with existing methods. Finally, we calculate the $T$-count to perform phase-estimation on Jellium. In the low-filling regime, we observe improvements in gate complexity of over $10\times$ compared to the best Trotter-based approach reported to date. We also report gate counts competitive with qubitization-based approaches for Wigner-Seitz values of physical interest.
\end{abstract}

\maketitle

\section{Introduction}\label{Sec:Intro}

There is considerable interest in whether quantum computers -- both those available at present, and those under development -- can be used to solve problems of scientific and commercial importance. This is particularly evident in the field of quantum simulation of chemical systems -- for recent reviews of progress in this area, we direct the reader to Refs.~\cite{mcardle2020review, cao2019review, bauer2020quantum}. Several algorithms have been developed to obtain the eigenstates of chemical systems. These include variational quantum algorithms~\cite{peruzzo2014variational,mcclean2016theory} that aim to maximise the limited coherence times of currently available hardware. However, this comes at the cost of introducing heuristic aspects, making it difficult to obtain rigorous performance guarantees. In contrast, approaches based on quantum phase estimation~\cite{Abrams1999PhaseEst, aspuru2005simulated} provide a route to calculate eigenstates to within a specifiable error, assuming only that we can efficiently prepare approximate eigenstates with sufficiently high overlap with the true eigenstates.

The resources we allocate to a fault tolerant quantum computation will depend on our ability to bound errors in the algorithm; the tighter our error estimates, the fewer resources we will require. Several previous works have estimated the resources required for phase estimation based on product-formula decompositions (also known as Trotterization)~\cite{wecker2014gates, wecker2015progress, poulin2014trotter, babbush2015chemical, reiher2017elucidating, motta2018low, Kivlichan2020improvedfault}. It was recently shown by Su, Huang and Campbell~\cite{su2020nearly} that knowledge about the number of fermions present in a chemical system can be exploited to improve the asymptotic performance of Trotterization. That work introduced an error metric, termed the \emph{fermionic seminorm}, to bound the Trotter error. This approach uses knowledge of the number of fermions in the system to offset the dependence of the error on the number of orbitals. 
This effect may be particularly important for applications to chemical systems in realistically sized basis sets, which will need to be large in order to accurately resolve dynamic correlation in the wavefunction.  The Su-Huang-Campbell (SHC) bound aimed to find an analytic bound with the best asymptotic complexity.  Here we present complementary work that also uses the fermionic seminorm with the goal of developing techniques for numerically obtaining bounds with best performance in practice.

In this work, we introduce three factorized decompositions of the electronic structure Hamiltonian in a plane wave dual basis, and use these in conjunction with the fermionic seminorm to obtain tighter Trotter error bounds in practice. Our approach is inspired by prior work using low-rank decompositions to reduce the number of terms in a Hamiltonian and thereby reduce the gate complexity of quantum algorithms~\cite{motta2018low,Berry2019qubitization,lee2020TensorHyperContraction,vonburg2020catalysis}. However, our use of factorized decompositions is purely computational and optimised for tightest error bounds, with no corresponding change in the execution of the quantum algorithm. A high-level overview of our approach can be found in \sec{ImprovedBounds}.

Each of our three factorized decompositions exhibits its own advantage.  The spectral decomposition is generally applicable and extends beyond the plane wave dual basis.  The cosine decomposition best exploits fermion number information and so performs the most effectively in the low-filling fraction regime.  The Cholesky decomposition has the smallest constant factor overhead and so performs best in the medium and half-filling regimes. We discuss these decompositions in detail and compare the resulting Trotter error bounds in \sec{DecompsAndBounds}.

These performance observations are supported by numerical results in \sec{NumericalResults}, obtained by applying our approach to the uniform electron gas (Jellium) introduced in \sec{Jellium}.  In these numerics, we also benchmark against three prior art bounds: the analytic SHC bound described earlier~\cite{su2020nearly}; the fermionic commutator approach used by Kivlichan \textit{et al}~\cite{Kivlichan2020improvedfault}; and a similar Pauli commutator approach where there is anecdotal evidence of good performance (see App.\ A of Ref.~\cite{huggins2019efficient}). We report a substantial classical runtime advantage for the calculation of our bounds. The fermionic and Pauli commutator approaches became intractable to calculate at larger spin-orbital number $N$, so could not be computed beyond $N \sim 200$, without access to $> 100$~GB of RAM. In contrast, it took fewer than 6 hours (using a 3.6GHz c5.2xlarge EC2 instance on AWS) to calculate our new bounds on a 512 spin-orbital instance, using $< 16$~GB of RAM.

One target problem for Trotter methods has been for phase estimation of the ground state energy of the uniform electron gas~\cite{Kivlichan2020improvedfault}.  Using our improved Trotter error bounds for Jellium, we calculate the $T$-count for this problem and demonstrate the expected improvements in runtime. We also compare our gate counts to those obtained using qubitization~\cite{babbush2018EncodingElectronicSpectra}, and find comparable results in some parameter regimes of interest.

We present mathematical preliminaries in \sec{MathBackground} that are necessary to understand our factorized decompositions and their numerical implementations. We conclude the paper in \sec{Discussion} with a brief summary of our contributions and a collection of avenues for future work.

\section{Preliminaries}\label{Sec:MathBackground}

\subsection{Fermionic systems and seminorm}

The electronic structure Hamiltonian is a widely used model for molecular and material systems where the positions of the nuclei are considered fixed.  In an arbitrary basis of $N$ electronic spin-orbitals, the Hamiltonian can be written as
\begin{equation}\label{eq:ArbBasisSecondQuantizedHamiltonian}
    H = \sum_{pq} h_{pq} a_p^\dag a_q + \sum_{pqrs} h_{pqrs} a_p^\dag a_q^\dag a_r a_s,
\end{equation}
where $a_s$ is the fermionic annihilation operator on spin-orbital $s$ and the coefficients $ h_{pq}$ and $h_{pqrs}$ are defined by integrals over the basis functions~\cite{helgaker2014molecular}.   Using the plane wave dual basis given by~\cite{babbush2018planewaves}, the number of terms is reduced from $\mathcal{O}(N^4)$ to $\mathcal{O}(N^2)$ with the simple form
\begin{equation} \label{eq:PlaneWaveDualHamiltonian}
    H = \sum_{p,q} T_{pq} a^\dag_p a_q + \sum_{p} U_p n_p + \sum_{p \neq q} V_{pq} n_p n_q,
\end{equation}
which is split into the electron kinetic, electron-nuclei, and electron-electron terms, respectively. The coefficients $T_{pq}$, $U_p$, and $V_{pq}$ are defined by integrals over the basis functions, as discussed in \sec{DecompsAndBounds}.

When simulating time evolution under a Hamiltonian (such as those given above), the error is typically quantified using the spectral-norm distance between the time evolution operator, and the quantum circuit used to approximate it. However, it is possible to use knowledge about the initial state to improve the error bound. In Ref.~\cite{su2020nearly} the fermionic seminorm of an operator $X$ was defined as the maximum transition amplitude of the operator between two states in the $\eta$-electron subspace
\begin{equation}\label{Eq:FSemiDef}
    ||X||_\eta := \mathrm{max}_{\ket{\psi_\eta}, \ket{\phi_\eta}} |\bra{\phi_\eta} X \ket{\psi_\eta}|.
\end{equation}
We say an operator $X$ is number preserving if $X$ acting on an $\eta$-electron state yields some other $\eta$-electron state. It was shown in Ref.~\cite{su2020nearly} that the fermionic seminorm has similar properties to well-known existing norms. For number preserving operators $X$, $Y$, we will make use of the following properties:
\begin{itemize}
    \item $||X + Y||_\eta \leq ||X||_\eta + ||Y||_\eta$ (Triangle inequality) 
    \item $||X \cdot Y||_\eta \leq ||X||_\eta \cdot ||Y||_\eta$ (H\"older inequality) 
    \item $||\lambda X ||_\eta = |\lambda| \cdot ||X||_\eta$ (for $\lambda \in \mathbb{C}$)
    \item $||X^\dag||_\eta = ||X||_\eta$
    \item $||UXW||_\eta = ||X||_\eta$ (for $U, X, W$ number preserving, $U, W$ unitary)
\end{itemize}
We remark that it is a seminorm rather than a norm because it can evaluate to zero for some non-zero operators. For example, for a system with a single fermion, we have $\norm{n_pn_q}_{\eta=1}=0$ (for $p\neq q$), but $n_pn_q$ is a nonzero operator.

\subsection{Prior art in commutator bounds}

This work considers Trotter-based approaches to implement the time evolution operator that is used in Hamiltonian simulation and quantum phase estimation. For a Hamiltonian that can be decomposed as $H = \sum_{j=1}^M H_j$, a first-order Trotter decomposition approximates the time evolution operator as
\begin{equation} \label{eq:FirstOrder}
    e^{iHt} \approx \prod_{j=1}^M e^{ it H_j}  =: U_1
\end{equation}
and a second-order Trotter decomposition approximates the time evolution operator as
\begin{equation} \label{eq:SecondOrder}
    e^{iHt} \approx \bigg{(}\prod_{j=1}^M e^{\frac{it}{2}H_j} \bigg{)} \bigg{(} \prod_{j=M}^1 e^{\frac{it}{2}H_j} \bigg{)} =: U_2. 
\end{equation}
It has been shown~\cite{Kivlichan2020improvedfault,childs2019trottererror} that this approximation has an error given by
\begin{align} 
 ||e^{iHt} - U_1 ||& \leq W_1  t^2, \\
    ||e^{iHt} - U_2 || & \leq  W_2 t^3 ,
\end{align}
where $W_1$ and $W_2$ are defined as
\begin{align} \label{CommutatorBound}
W_1 & :=  \frac{1}{2} \sum_{a=1}^{M}  \bigg{|}\bigg{|} \sum_{b > a}^M  [H_b, H_a] \bigg{|}\bigg{|} , \\ 
W_2 & :=  \frac{1}{12} \sum_{a=1}^{M} \bigg{(} \bigg{|}\bigg{|} \sum_{c > a} \sum_{b > a} [H_c, [H_b, H_a]] \bigg{|}\bigg{|}
    + \frac{1}{2}\bigg{|}\bigg{|}\sum_{b>a} [H_a, [H_a, H_b]] \bigg{|}\bigg{|} \bigg{)},
\end{align}
where $||...||$ denotes the operator norm (also known as the spectral norm -- i.e. the largest singular value of the operator).  

In practice, it can be difficult to get a tight value for $W_{1/2}$ because of the complexity in evaluating the operator norm of a high-dimensional operator such as $[H_c, [H_b, H_a]]$. As such, in aid of numerical expediency, a further relaxation is often made.  Each nested commutator is expanded in terms of operators $P_j$ with known operator norm $||P_j||=1$ so that
\begin{equation}
    \sum_{c>a} \sum_{b>a} [H_c, [H_b, H_a]] = \sum_{j} \alpha_j P_j
\end{equation}
and then one can bound
\begin{equation}
   \bigg{|}\bigg{|}  \sum_{c>a} \sum_{b>a} [H_c, [H_b, H_a]] \bigg{|}\bigg{|} \leq \sum_{j} |\alpha_j| .
\end{equation}
Common choices include choosing $P_j$ as tensor products of Pauli operators, or as fermionic excitation operators (e.g. $P_{j}=a_{j_1}^\dagger a_{j_2}^\dagger a_{j_3}^\dagger a_{j_4} a_{j_5} a_{j_6}$). Throughout, we refer to bounds using these relaxations as the Pauli commutator bound and Fermionic commutator bound, respectively. For example, Ref.~\cite{Kivlichan2020improvedfault} used the Fermionic commutator bound to estimate the resources for phase estimation in the plane wave dual basis. In our numerical examples, we will benchmark against these prior art bounds.

Ref.~\cite{su2020nearly} showed the commutator bounds can be tightened in the special case where $H$ is a fermionic Hamiltonian and every $H_j$ in the Trotter decomposition (\eq{SecondOrder}) is number-preserving, so that for second-order Trotter 
\begin{align}\label{Eq:TrotterBoundsGeneralSeminorm}
    W_2 \leq \frac{1}{12} \sum_{a=1}^{M} \bigg{(} \bigg{|}\bigg{|} \sum_{c > a} \sum_{b > a} [H_c, [H_b, H_a]] \bigg{|}\bigg{|}_\eta
    + \frac{1}{2}\bigg{|}\bigg{|}\sum_{b>a} [H_a, [H_a, H_b]] \bigg{|}\bigg{|}_\eta \bigg{)}.
\end{align}
where the operator norm has been replaced by the tighter fermionic semi-norm.  Ref.~\cite{su2020nearly} further considered Hamiltonians in the plane wave dual basis (recall \eq{PlaneWaveDualHamiltonian}) and a Trotterization where the Hamiltonian is considered as containing two terms; $H_t = \sum_{pq} T_{pq} a^\dag_p a_q$ and $H_v=
\sum_{p q} \bar{V}_{pq} n_p n_q$. Ref.~\cite{su2020nearly} derived bounds for arbitrary order product formulae, with the second-order result
\begin{align}\label{Eq:SHC_asymptotic}
    W_2 & \leq \mathcal{O}( ||\bar{V}||_{\mathrm{max}}^2 ||T||\eta^3    + ||T||^2 ||\bar{V}||_{\mathrm{max}}\eta^2   ),
\end{align}
where $|| \ldots ||_{\mathrm{max}}$ is the max-norm that represents the largest matrix element in absolute value.  A key observation is that the bound depends on $\eta$ and so captures the expected dependence on the fermion number. The big-$\mathcal{O}$ of this result hides the constant factors that are needed for numerical comparisons. For the case of the plane wave dual basis, \Eq{TrotterBoundsGeneralSeminorm} reduces to a sum of two terms. In \append{AnalyticBounds} we have evaluated these terms, which are given by
\begin{align}
        \big{|}\big{|} [[H_t, H_v], H_t] \big{|}\big{|}_\eta &\leq 4 \cdot \big{|}\big{|} T \big{|}\big{|}^2 \cdot \big{|}\big{|} \bar{V} \big{|}\big{|}_{\mathrm{max}} \cdot \eta \cdot (4 \eta + 1),\\
        \big{|}\big{|} [[H_t, H_v], H_v] \big{|}\big{|}_\eta &\leq 12 \cdot \big{|}\big{|} T \big{|}\big{|} \cdot \big{|}\big{|} \bar{V} \big{|}\big{|}_{\mathrm{max}}^2 \cdot \eta^2 \cdot (2 \eta + 1).
\end{align}
We refer to this as the `SHC bound' throughout.

\section{Improved fermionic seminorm bounds}\label{Sec:ImprovedBounds}

In this work, we make particular use of the properties of free-fermionic Hamiltonians $H(A) := \sum_{i,j} A_{ij} a_i^\dag a_j$. We refer to $A$ as the coefficient matrix of the free-fermionic Hamiltonian. A free-fermionic Hamiltonian can be efficiently diagonalised by diagonalising its coefficient matrix. We can then calculate the fermionic seminorm of a free-fermionic Hamiltonian as
\begin{equation}
    \begin{aligned}
        ||H(A)||_\eta &= \big{|}\big{|}\sum_{i,j} A_{ij} a_i^\dag a_j \big{|}\big{|}_\eta \\
        &= \big{|}\big{|}V \big{(}\sum_{i,j} A_{ij} a_i^\dag a_j \big{)} V^{-1} \big{|}\big{|}_\eta \\
        &= \big{|}\big{|}\sum_{k} \lambda_k \tilde{a}_k^\dag \tilde{a}_k\big{|}\big{|}_\eta,
    \end{aligned}
\end{equation}
where $V$ is a unitary matrix that diagonalises the free-fermionic Hamiltonian, and $\lambda_k$ are the eigenvalues of the coefficient matrix $A$. This expression can be evaluated using Eq.~(\ref{Eq:FSemiDef}) to give
\begin{equation} \label{Eq:HT}
    || H(A) ||_{\eta} = \sfs{A} := \mathrm{max} \left\{ |  \sum_{\lambda \in S} \lambda| ; |S|=\eta , S \subseteq \lambda(A) \right\} .
\end{equation}
Here we have defined another seminorm $\sfs{\cdot}$ which takes a coefficient matrix $A$ as its argument. We call this the \textit{reduced} fermionic semi-norm as the argument is a smaller $N$-by-$N$  matrix $A$, rather than the large operator $H(A)$ that is represented by a $2^N$-by-$2^N$ matrix. The result of \Eq{HT} tells us that for free-fermionic operators $H(A)$ the problem of evaluating the fermionic semi-norm simplifies to the easier problem of evaluating the reduced fermionic seminorm.  Evaluating the reduced fermionic semi-norm takes the set $\lambda(A)$ of eigenvalues of $A$ and finds the subset $S \subset \lambda(A)$ with $\eta$ elements and largest sum in absolute value. If $A$ is Hermitian this is further simplified, as we can consider the sum of the $\eta$ largest eigenvalues, and the sum of the $\eta$ most-negative eigenvalues, and choose the larger absolute value.  Therefore, \Eq{HT} can be efficiently computed for Hermitian $A$.

Another useful property involves the commutator of two free-fermionic Hamiltonians
\begin{equation}
    [H(A), H(B)] = H([A,B]),
\end{equation}
itself a free-fermionic Hamiltonian. This has previously been noted and made use of in the context of quantum simulation in Refs.~\cite{poulin2014trotter, campbell2020early}.

Motivated by these properties of free-fermionic Hamiltonians, we consider decomposing the Hamiltonian as
\begin{equation}
  H = H(A) + \sum_l H(X\defindex{l}) H(Y\defindex{l}),
\end{equation}
where $A, X\defindex{l}, Y\defindex{l}$ are $N \times N$ coefficient matrices, and $N$ is the number of spin-orbitals considered. Decompositions of this form have been considered in the context of quantum computing in Refs.~\cite{poulin2014trotter, motta2018low}, where they were obtained by eigen/Cholesky decompositions of the tensor $h_{pqrs}$. This yields the Hamiltonian in a `single factorised' form~\cite{motta2018low, Berry2019qubitization, huggins2019efficient}. For example, if we consider the electronic structure Hamiltonian in a Gaussian orbital basis set (described by \eq{ArbBasisSecondQuantizedHamiltonian}) we can apply a spectral decomposition of the tensor $h_{pqrs}$ to write the Hamiltonian in the form (see \append{Spectral})
\begin{equation} \label{Eq:GeneralDecomp}
    H = H(\tilde{h}) + \sum_{\ell=1}^L \lambda_{\ell} H(X\defindex{\ell}) H(X\defindex{\ell})
\end{equation}
where $L$ denotes the number of terms in the spectral decomposition, and $\lambda_{\ell}$ are the corresponding eigenvalues. We consider Trotter decompositions with each term $H_j$ in \eq{FirstOrder} or \eq{SecondOrder} corresponding to some subset of terms from \Eq{GeneralDecomp}. We show in \append{Spectral} that we can bound the first-order Trotter error with the commutator bound
\begin{align} \label{eq:SpectralBoundMainText}
    W_1 \leq \sum_{j=1}^L \bigg{(} |\lambda_j | \cdot \sfs{ [\tilde{h}, X\defindex{j}] } \cdot \sfs{X\defindex{j}} \bigg{)} + 2 \sum_{i=1, j > i}^L \bigg{(} |\lambda_i| \cdot |\lambda_j| \cdot \sfs{[X\defindex{i}, X\defindex{j}]} \cdot \sfs{X\defindex{i}} \cdot \sfs{X\defindex{j}} \bigg{)}
\end{align}
To obtain this form, we made use of commutator identities such as $[A, BC] = [A,B]C + B[A,C]$ and $[AB, CD] = A[B,C]D + CA[B,D] + [A,C]BD + C[A,D]B$. Similar bounds can be obtained for higher-order Trotter formulae, and in \append{SecondOrderDecomp} we present second-order bounds for the special case of the plane wave dual Hamiltonian.

A similar approach was attempted in Ref.~\cite{poulin2014trotter}, however, that work did not explicitly make use of information about the number of electrons in the system, and therefore the result is not tight in the low-filling regime. Depending on the form of $h_{pqrs}$, other decompositions may be possible. In the following section, we present three decompositions of the plane wave dual basis Hamiltonian, motivated by its simple form, and the analytic expressions available for the Hamiltonian coefficients in this basis. We summarise the main features of these decompositions in \tab{BoundComparison}. Each of these three decompositions has a particular benefit; the spectral decomposition is the extension of the approach discussed above (and in \append{Spectral}) to the plane wave dual basis, and so is generally applicable to any orbital basis. The Cholesky decomposition performs best in the half-filling regime ($\eta=N/2$), while the cosine decomposition performs best in the low-filling regime ($\eta \ll N/2$). All of these bounds are more efficient to compute than the existing fermionic and Pauli commutator bounds. \\

\begin{table*}[]
    \centering
    \begin{tabular}{c|c|c|c|c|c}
         Approach & Memory & Runtime & Exploits   & $\eta \ll N/2$ & $\eta = N/2$  \\
      &  &  &  fermion \# &  rank &  rank \\
         \hline
         Fermionic commutator~\cite{Kivlichan2020improvedfault} & $\mathcal{O}(N^4)$ & $\mathcal{O}(N^6)$ & No & 5th/5 & 2nd/5 \\
         Pauli commutator & $\mathcal{O}(N^5)$ & $\mathcal{O}(N^6)$ & No & -- & -- \\
         SHC bound \cite{su2020nearly}  & $\mathcal{O}(N^2)$ & $\mathcal{O}(N^3)$ & Yes & 3rd/5 & 5th/5 \\
        Spectral decomp. [This work] & $\mathcal{O}(N^2)$ & $\mathcal{O}(N^5)$ & Yes & 4th/5 & 4th/5 \\
        Cholesky decomp. [This work] & $\mathcal{O}(N^2)$ & $\mathcal{O}(N^5)$ & Partially & 2nd/5 & 1st/5 \\
         Cosine decomp. [This work] & $\mathcal{O}(N^2)$ & $\mathcal{O}(N^5)$ & Yes & 1st/5 & 3rd/5
    \end{tabular}
\caption{A comparison of the different Trotter error bounds considered. The memory and runtime scaling are given for calculations of the second-order bounds, as outlined in \append{ComputationDetails}. The final two columns rank the second-order data presented in \fig{FixedOrbs}, for a uniform electron gas system with 200 spin-orbitals, and varying electron number. The memory requirement of the Pauli commutator approach was too severe to carry out this second-order Trotter calculation, though  we will later present first-order Trotter results for this approach.}
    \label{tab:BoundComparison}
\end{table*}

\section{Plane wave dual basis decompositions and Trotter error bounds}\label{Sec:DecompsAndBounds}
The plane wave dual basis electronic structure Hamiltonian given by \eq{PlaneWaveDualHamiltonian} describes a system with $\eta$ electrons in a simulation box of size $\Omega \propto L^d$, where $d$ is the dimensionality of the system, and $L$ is the number of grid points along each side of the box. The spin-orbitals are obtained from a discrete Fourier transform of plane waves. These plane waves are defined by
\begin{equation}
    \begin{aligned}
        \phi_{\vec{\nu}}(\vec{r}) = \sqrt{\frac{1}{\Omega}} e^{i \vec{k}_{\vec{\nu}} \cdot \vec{r}} \quad \quad \quad
        \vec{k}_{\vec{\nu}} = \frac{2\pi \vec{\nu}}{\Omega^{1/d}} \quad \quad \quad
        \vec{\nu} &\in \bigg{[} -\left \lfloor{\frac{L}{2}} \right \rfloor, \left \lfloor{\frac{L}{2}} \right \rfloor \bigg{)}^d \in \mathbb{Z}^d,
    \end{aligned}
\end{equation}
where $N$ is the number of spin-orbital basis functions used, and $\vec{\nu}$ enumerates the $N/2$ possible distinct momentum vectors of the system. Note that if $L$ is odd, the interval of $\vec{\nu}$ is closed, rather than half-open. The plane wave dual basis resembles a smooth approximation to a grid of delta functions. The coefficients in \eq{PlaneWaveDualHamiltonian} are given by~\cite{babbush2018planewaves}
\begin{equation}\label{Eq:Coefficients}
    \begin{aligned}
    T_{pq} = \delta_{\sigma_p, \sigma_q} \sum_{\vec{\nu}} &\frac{\vec{k}_{\vec{\nu}}^2 \mathrm{cos}(\vec{k}_{\vec{\nu}} \cdot (\vec{r}_{\vec{p}} - \vec{r}_{\vec{q}}))}{N} \\
    U_{p} = - \sum_{j, \vec{\nu}: |\vec{\nu}| \neq 0} \frac{4\pi \zeta_j \mathrm{cos}(\vec{k}_{\vec{\nu}} \cdot (\vec{R}_{j} - \vec{r}_{\vec{p}}))}{\Omega \vec{k}_{\vec{\nu}}^2} \quad &\quad \quad
    V_{pq} = \sum_{\vec{\nu}: |\vec{\nu}| \neq 0} \frac{2\pi \mathrm{cos}(\vec{k}_{\vec{\nu}} \cdot (\vec{r}_{\vec{p}} - \vec{r}_{\vec{q}}))}{\Omega \vec{k}_{\vec{\nu}}^2}.
    \end{aligned}
\end{equation}
Here, $\vec{r}_{\vec{p}}$ is the position of the orbital centroid corresponding to spatial-orbital $\vec{p}$
\begin{equation}
   \vec{r}_{\vec{p}} = \vec{p}\bigg{(}\frac{2\Omega}{N}\bigg{)}^{\frac{1}{d}}  \quad \quad \quad  \vec{p} \in \bigg{[} -\left \lfloor{\frac{L}{2}} \right \rfloor, \left \lfloor{\frac{L}{2}} \right \rfloor \bigg{)}^d \in \mathbb{Z}^d,
\end{equation}
$\sigma_p$ is the spin of the $p$th spin-orbital (here, we have mapped the vector index $\vec{p}$ to an integer value by defining an ordering for the spin-orbital basis functions), and $\vec{R}_j$ and $\zeta_j$ are the position and charge of the $j$th nucleus in the system.

In the plane wave dual basis, the Hamiltonian terms can be partitioned into kinetic and potential terms, respectively
\begin{align}
    H_t = \sum_{p,q} T_{pq} a_p^\dag a_q \quad \quad \quad \quad     H_v = \sum_p U_p n_p + \sum_{p \neq q} V_{pq} n_p n_q.
\end{align}
We can approximate the time evolution operator by applying the potential terms (which all commute with each other, and so induce no Trotter error), implementing a basis change to plane waves, such that the kinetic term becomes diagonal and can be implemented without Trotter error, and then changing back to the plane wave dual basis (or the equivalent, but starting in the plane wave basis). The second-order Trotter error for these approaches are given by
\begin{equation}
\begin{aligned}
	\norm{e^{itH} - e^{i\frac{t}{2}H_v}e^{itH_t}e^{i\frac{t}{2}H_v}}_\eta &\leq \frac{t^3}{12} \bigg{(} \big{|}\big{|} [[H_t, H_v], H_t] \big{|}\big{|}_\eta + \frac{1}{2}\big{|}\big{|} [[H_t, H_v], H_v] \big{|}\big{|}_\eta \bigg{)} \\
	\norm{e^{itH} - e^{i\frac{t}{2}H_t}e^{itH_v}e^{i\frac{t}{2}H_t}}_\eta &\leq \frac{t^3}{12} \bigg{(} \big{|}\big{|} [[H_t, H_v], H_v] \big{|}\big{|}_\eta + \frac{1}{2}\big{|}\big{|} [[H_t, H_v], H_t] \big{|}\big{|}_\eta \bigg{)}	
\end{aligned}
\end{equation}
The kinetic and electron-nuclei interaction terms are free-fermionic Hamiltonians. This section presents three ways to decompose the electron-electron interaction term into a sum of products of free-fermionic Hamiltonians such that we can write $H_v = H(U) + \sum_l H(X\defindex{l}) H(Y\defindex{l})$. We use these decompositions, the aforementioned commutator identities, and the fermionic seminorm properties of free-fermion Hamiltonians to derive expressions for first- and second-order commutator bounds.

We calculate the first-order bound here, and refer the reader to \append{SecondOrderDecomp} for calculations of the second-order bounds. The first-order commutator is given by
\begin{equation}
    \begin{aligned}
        [H_t, H_v] &= [H(T), H(U) + \sum_l H(X\defindex{l}) H(Y\defindex{l})] \\
        &= [H(T), H(U)] + \sum_l [H(T), H(X\defindex{l}) H(Y\defindex{l})] \\
    \end{aligned}
    \end{equation}
We can simplify the second term using $[A, BC] = [A,B]C + B[A,C]$ to give
\begin{equation}
\begin{aligned}
        [H_t, H_v] &= H([T,U]) + \sum_l [H(T), H(X\defindex{l})] H(Y\defindex{l}) +  H(X\defindex{l}) [H(T), H(Y\defindex{l})] \\
        &= H([T, U]) + \sum_l H([T, X\defindex{l}]) H(Y\defindex{l}) + H(X\defindex{l}) H([T, Y\defindex{l}]).
    \end{aligned}
\end{equation}
Using the triangle and H\"older inequalities, the fermionic seminorm of the first-order commutator is then upper bounded by
\begin{align}\label{Eq:KeyFirstOrderBound}
       \big{|}\big{|} [H_t, H_v] \big{|}\big{|}_\eta & \leq \sfs{[T,U]} + \sum_l \bigg{(} \sfs{[T, X\defindex{l}]} \cdot \sfs{Y\defindex{l}} + \sfs{[T, Y\defindex{l}]} \cdot \sfs{X\defindex{l}} \bigg{)}.
\end{align}

\subsection{Chemical potentials}\label{Subsec:ChemicalPotential}
When working in a fixed particle number manifold, we can shift the chemical potential of the problem to try and reduce the resulting Trotter error bound. This technique has previously been found to be beneficial in simulations of the Fermi-Hubbard model~\cite{campbell2020early}. We can transform the Hamiltonian to
\begin{align}
    H_v &\rightarrow H_v + C \eta \nonumber \\
    &= H(U) + \sum_{p \neq q} V_{pq} n_p n_q + C \sum_{p} n_p \nonumber \\
    &= H(U) + \sum_{p q} (\delta_{pq}C+V_{pq}) n_p n_q
\end{align}
where we have used that $n_p^2 = n_p$. This transformation adds a constant $C$ to the diagonal of $V_{pq}$.

\subsection{Spectral decomposition}\label{Subsec:SpectralDecomp}
In the plane wave dual basis, the electron-electron Coulomb interaction matrix $V_{pq}$ is real symmetric, and therefore admits a spectral decomposition 
\begin{equation}
    V_{pq} = \sum_i \lambda_i [v\defindex{i}]_p [v\defindex{i}]_q.
\end{equation}
While Eq.~(\ref{eq:PlaneWaveDualHamiltonian}) corresponds to defining $V$ with $V_{pp} := 0$, we can also use the chemical potential shift outlined above to set $V_{pp} := C$. We factorise the Hamiltonian as
\begin{equation}
    \begin{aligned}
    H_v &= H(U) + \sum_{p,q} V_{pq} n_p n_q \\
    &= H(U) + \sum_{p,q, i} \lambda_i [v\defindex{i}]_p [v\defindex{i}]_q n_p n_q \\
    &= H(U) + \sum_{i} \lambda_i \bigg{(} \sum_p [v\defindex{i}]_p n_p \bigg{)} \bigg{(} \sum_q [v\defindex{i}]_q n_q \bigg{)} \\
    &:= H(U) + \sum_{i} \lambda_i H(v\defindex{i}) H(v\defindex{i})
    \end{aligned}
\end{equation}
Here, $v\defindex{i}$ are diagonal $N \times N$ coefficient matrices. The first-order bound is given by \begin{equation}
\begin{aligned}
    \big{|} \big{|} [H_t, H_v] \big{|} \big{|}_\eta &\leq \sfs{[T,U]} + 2 \sum_{i} |\lambda_i| \bigg{(} \sfs{[T, v\defindex{i}]} \cdot \sfs{v\defindex{i}} \bigg{)}
\end{aligned}
\end{equation}
The second-order bounds are given in \append{SpecificSecondOrderBounds}.  This decomposition can be regarded as an instance of the general approach of spectral decomposing tensors $h_{pqrs}$ and we discuss this further in \append{Spectral}.
\\

\subsection{Cholesky decomposition}\label{Subsec:Cholesky}
We can also consider a Cholesky decomposition of the matrix $V$. The Cholesky decomposition factorises a positive (semi)-definite Hermitian matrix into the product of a lower triangular matrix and its Hermitian conjugate, $V = LL^\dag$. For the real symmetric matrix $V_{pq}$, we first shift the chemical potential to make $V$ positive definite. The Cholesky decomposition is then given by
\begin{equation}
    V_{pq} = \sum_{i} L_{pi}L_{iq}^T.
\end{equation}
We can then factorise the Hamiltonian as
\begin{align}
    H_v &= H(U) + \sum_{pq} V_{pq} n_p n_q \nonumber \\
    &= H(U) + \sum_{ipq} L_{pi}L_{iq}^T n_p n_q \nonumber \\
    &= H(U) + \sum_i \bigg{(} \sum_{p} L_{pi} n_p \bigg{)} \bigg{(} \sum_{q} L_{qi} n_q \bigg{)} \nonumber \\
    &:= H(U) + \sum_i H(L\defindex{i}) H(L\defindex{i})
\end{align}
where $L\defindex{i}$ are diagonal coefficient matrices such that $[L\defindex{i}]_{pq} = \delta_{pq} L_{qi}$. The first-order bound is given by \begin{equation}
\begin{aligned}
    \big{|} \big{|} [H_t, H_v] \big{|} \big{|}_\eta &\leq \sfs{[T,U]} + 2 \sum_{i} \bigg{(} \sfs{[T, L\defindex{i}]} \cdot \sfs{L\defindex{i}} \bigg{)}
\end{aligned}
\end{equation}
The second-order bounds are given in \append{SpecificSecondOrderBounds}. As the Cholesky matrix $L$ is lower triangular, the free-fermionic Hamiltonians $H(L\defindex{i})$ become increasingly low rank at higher values of $i$, suggesting that this decomposition may not fully exploit fermion number.

\subsection{Cosine decomposition}\label{Subsec:CosineDecomp}
We consider the following decomposition that depends explicitly on the structure of the terms in the matrix $V_{pq}$.
We introduce the shorthand $\omega_\nu^p := \vec{k}_\nu \cdot \vec{r}_p$. Applying the double angle formula to \Eq{Coefficients} yields
\begin{equation}
    \begin{aligned}
    V_{pq} &= \frac{2\pi}{\Omega} \sum_{\nu \neq 0} \frac{1}{|\vec{k}_\nu|^2} \mathrm{cos}(\omega_\nu^p - \omega_\nu^q) \\
   &= \frac{2\pi}{\Omega} \sum_{\nu \neq 0} \frac{1}{|\vec{k}_\nu|^2} \bigg{(} \mathrm{cos}(\omega_\nu^p)\mathrm{cos}(\omega_\nu^q) + \mathrm{sin}(\omega_\nu^p)\mathrm{sin}(\omega_\nu^q) \bigg{)}.
    \end{aligned}
\end{equation}
We can use this to write
\begin{equation}\label{Eq:Decomp1}
    \begin{aligned}
    \sum_{p \neq q} V_{pq} n_p n_q 
   = & \frac{2\pi}{\Omega} \sum_{p} \sum_{q} \sum_{\nu \neq 0} \frac{1}{|\vec{k}_\nu|^2} \bigg{(} \mathrm{cos}(\omega_\nu^p) n_p  \mathrm{cos}(\omega_\nu^q)n_q + \mathrm{sin}(\omega_\nu^p)n_p  \mathrm{sin}(\omega_\nu^q) n_q \bigg{)} \\ 
    & - \frac{2\pi}{\Omega} \sum_{p} \sum_{\nu \neq 0} \frac{1}{|\vec{k}_\nu|^2} \bigg{(} \mathrm{cos}^2(\omega_\nu^p) n_p + \mathrm{sin}^2(\omega_\nu^p)n_p \bigg{)} \\
    =& \frac{2\pi}{\Omega} \sum_{\nu \neq 0} \frac{1}{|\vec{k}_\nu|^2} \bigg{(}\sum_{p} \mathrm{cos}(\omega_\nu^p) n_p\bigg{)}  \bigg{(}\sum_{q} \mathrm{cos}(\omega_\nu^q) n_q\bigg{)}  \\ 
    & + \frac{2\pi}{\Omega} \sum_{\nu \neq 0} \frac{1}{|\vec{k}_\nu|^2}  \bigg{(}\sum_{p} \mathrm{sin}(\omega_\nu^p) n_p\bigg{)}  \bigg{(}\sum_{q} \mathrm{sin}(\omega_\nu^q) n_q\bigg{)} - \frac{2\pi}{\Omega} \sum_{p} \sum_{\nu \neq 0} \frac{1}{|\vec{k}_\nu|^2} n_p
    \end{aligned}
\end{equation}
In the fixed electron-number manifold, the final term will only contribute a global phase during Hamiltonian simulation, and so can be dropped. We can rewrite $H_v$ as
\begin{equation}
    \begin{aligned}
    H_v &= H(U) + \sum_{\nu \neq 0} H(C\defindex{\nu}) H(C\defindex{\nu}) + H(S\defindex{\nu})H(S\defindex{\nu}),
    \end{aligned}
\end{equation}
where $C\defindex{\nu}$ and $S\defindex{\nu}$ are diagonal $N \times N$ coefficient matrices defined by
\begin{equation}
    \begin{aligned}
    [C\defindex{\nu}]_{ii} :=
        \sqrt{\frac{2\pi}{\Omega}} \frac{1}{|\vec{k}_\nu|} \mathrm{cos}(\omega_\nu^i)
    \quad \quad \quad \quad \quad
        [S\defindex{\nu}]_{ii} :=
        \sqrt{\frac{2\pi}{\Omega}} \frac{1}{|\vec{k}_\nu|} \mathrm{sin}(\omega_\nu^i)
    \end{aligned}
\end{equation}
The first-order bound is given by 
\begin{equation}
\begin{aligned}
    \big{|}\big{|} [H_t, H_v] \big{|}\big{|}_\eta &\leq \sfs{[T, U]} + 2 \sum_{\nu \neq 0} \sum_{A \in \{C, S\}} \bigg{(} \sfs{[T, A\defindex{\nu}]} \cdot \sfs{A\defindex{\nu}}  \bigg{)}
\end{aligned}
\end{equation}
The second-order bounds are given in \append{SpecificSecondOrderBounds}. We remark that it is possible to further simplify \Eq{Decomp1} to \begin{equation}
    \frac{2\pi}{\Omega} \sum_{\nu \neq 0} \frac{1}{|\vec{k}_\nu|^2} \bigg{[}\bigg{(}\sum_{p} e^{i\omega_\nu^p} n_p\bigg{)} \otimes \bigg{(}\sum_{q} e^{-i\omega_\nu^q} n_q\bigg{)} \bigg{]} - \frac{2\pi}{\Omega} \sum_{p} \sum_{\nu \neq 0} \frac{1}{|\vec{k}_\nu|^2} n_p.
\end{equation}
The more compact form of this decomposition suggests that it may offer a tighter bound. However, the resulting free-fermionic Hamiltonians $\sum_{p} e^{ \pm i\omega_\nu^p} n_p$ are non-Hermitian, and so yield operators that are neither Hermitian nor anti-Hermitian when commuted with the Hermitian kinetic operator. The resulting matrices may not be diagonalisable, making it unclear how to efficiently evaluate the fermionic seminorm of the operator. In \tab{BoundComparison}, we reported that the cosine decomposition is the top-ranked approach in the low-filling fraction regime. \\

\subsection{Outlook}

In the following sections, we will apply these bounds to the 2D uniform electron gas in a plane wave dual basis set. The Hamiltonian for this system is given by \eq{PlaneWaveDualHamiltonian}, but with $U_p = 0~\forall p$. As a result, all of the commutators containing $U$ can be dropped from the above expressions when considering this system. In the following section, we provide further background on the uniform electron gas. We then present numerical results comparing the Trotter error bounds derived above for the uniform electron gas, which form the basis of the rankings assigned in \tab{BoundComparison}.

\section{Uniform electron gas}\label{Sec:Jellium}
The uniform electron gas consists of $\eta$ electrons in a box of size $\Omega$. We are interested in the properties of this system as it scales to the thermodynamic limit -- where $\eta, \Omega \rightarrow \infty$, but the electron density $\rho = \eta/\Omega$ stays constant. At zero temperature, the physics of the system depends only on $\rho$. It is conventional to define a quantity referred to as the Wigner-Seitz radius $r_s$, that represents the average distance between electrons in the simulation cell. For a 3D simulation cell, the Wigner-Seitz radius is given by $r_s = (3/4\pi \rho)^{1/3}$ (for a 2D simulation cell, $r_s = \sqrt{1/\pi \rho}$). In order to make the system charge neutral, the electrons are immersed in a uniformly distributed sea of positive charge. Consequently, the system is often also referred to as `Jellium'. The Hamiltonian of the Jellium is given by~\cite{giuliani2005quantum}
\begin{align}
        H = \sum_i -\frac{\nabla_i^2}{2m} + \sum_{i < j} \frac{e^2}{|\vec{r}_i - \vec{r}_j|}  - \frac{e^2 \eta}{\Omega} \iint \sum_i^\eta \frac{\delta(\vec{r} - \vec{r}_i)}{| \vec{r} - \vec{r'} |} \mathrm{d}\vec{r} \mathrm{d}\vec{r'} + \frac{e^2 \eta^2}{2\Omega^2} \iint  \frac{1}{| \vec{r} - \vec{r'} |} \mathrm{d}\vec{r} \mathrm{d}\vec{r'}
\end{align}
where the first term represents the kinetic energy of the electrons, the second term describes the Coulomb repulsion of the electrons, the third term is interaction of the electrons with the uniform charge density of the positive background, and the final term is the self-interaction of the background charge. The long range nature of the Coulomb interaction causes divergences in the final two terms as the system scales to the thermodynamic limit. These divergences can be cancelled with a divergence of the opposite sign that arises in the electron-electron interaction term. The length scales are typically rescaled to be measured in Bohr radii ($\hbar^2/me^2$). When performing calculations on Jellium, we can either consider the real-space formulation of the problem discussed above, or project the Hamiltonian onto a basis set.

In addition to acting as a simple model of interacting electrons, the energy density of Jellium is used to parameterize some of the functionals used in density functional theory~\cite{kohn1965dft, perdew1992ParametrizationDFT, sun2010DPIparam}. Although the behaviour of Jellium is well understood in the low~\cite{wigner1934electrons} and high~\cite{fermi1926quantelung,bloch1929quantenmechanik,giuliani2005quantum} density limits, small energy differences in the intermediate regime lead to difficulty in resolving competing phases. This has led to unresolved questions about the existence of a superconducting phase in 2D Jellium~\cite{phillips1998superconductivity,ren1994jellium2DsuperconductorAnyon, takada1993jellium2Dsuperconducting}, as well as disagreements on the order of 0.7~mHartree per electron between different density functional parametrizations at electron densities of interest~\cite{ruggeri2018FCIQMCspinpolarisedJellium}. While existing computational techniques, such as quantum Monte Carlo methods, are able to obtain accurate energies of relatively large system sizes, these methods typically introduce an uncontrolled bias.
It is conventional to perform calculations on a succession of system sizes, which enables extrapolation to the thermodynamic limit. Extrapolation and correction for finite size effects~\cite{lin2001TwistAveraged, drummond2008FiniteSizeEffects, spink2013QMC3dJelliumSpinPolarised, ruggeri2018FCIQMCspinpolarisedJellium} 
often accounts for a large amount of the uncertainty present in the values estimated~\cite{giuliani2005quantum}.

Quantum Monte Carlo (QMC) methods, in particular, variational Monte Carlo (VMC) and diffusion Monte Carlo (DMC), are the leading techniques for calculating the ground state energy of Jellium. Following the pioneering calculations of Ceperley and Alder~\cite{ceperley1978QMC,ceperley1980FirstHEGQMC}, there have been a number of VMC/DMC calculations on both 3D Jellium~\cite{kwon1998Jellium3D, lopez2006jelliumbackflow, spink2013QMC3dJelliumSpinPolarised, ruggeri2018FCIQMCspinpolarisedJellium} and 2D Jellium~\cite{kwon1993Jellium2D, varsano2001spin, senatore2001spin, attaccalite2002jellium2DmonteCarlo, drummond2009PhaseDiagram2DHEG} (see Ref.~\cite{loos2016uniform} for a review of QMC calculations). Both VMC and DMC are typically performed in real-space, and have been applied to systems with on the order of $10^3$ electrons~\cite{giuliani2005quantum}. However, these methods are particularly susceptible to the fermion sign problem. This is typically mitigated by fixing the nodal points of the wavefunction to those of the trial wavefunction. Although this fixed node approximation is believed to work well for the uniform electron gas~\cite{senatore2001spin}, it introduces an uncontrolled bias that is not systematically improvable. While techniques can be used to mitigate this error, DMC energies for high density ($r_s \leq 5 $) electron gases are thought to possess an error of around 1~mHartree per electron (the fixed node error is believed to be smaller at larger $r_s$ values)~\cite{shepherd2012investigation, ruggeri2018FCIQMCspinpolarisedJellium}. State-of-the-art DMC calculations require on the order of $10^2$ CPU core hours~\cite{shepherd2012FCIQMCjellium}.

Calculations have also been performed using full configuration interaction quantum Monte Carlo (FCIQMC)~\cite{booth2009FCIQMC}, which evolves a population of random walkers using update rules that effectively propagate the wavefunction in imaginary time. FCIQMC is applied to systems that have been projected onto a basis set (typically plane waves for Jellium calculations). While this projection appears to mitigate the fermionic sign problem, it introduces a basis set error that must be eliminated by extrapolation to the continuum limit~\cite{shepherd2012ConvergenceBasisSet}. The basis set error decays as $1/N$, although this may be improved using explicitly correlated methods~\cite{luo2018TC_FCIQMC_planewave}. FCIQMC formally scales exponentially with the system size, but can in practice achieve bias-free results for small, weakly correlated Jellium systems (e.g. 19 electrons at $r_s=1$~\cite{ruggeri2018FCIQMCspinpolarisedJellium}). The approach is also practical for larger system sizes at high densities; producing more accurate results than DMC in 54 electron systems with $r_s \leq 1$~\cite{shepherd2012FCIQMCjellium}. Modern FCIQMC methods require around $10^3-10^5$ CPU core hours (depending on the value of $r_s$ investigated)~\cite{shepherd2012investigation, shepherd2012FCIQMCjellium}. As $r_s$ increases, the correlation present in the system becomes large, which makes FCIQMC methods too costly to converge~\cite{shepherd2012investigation, shepherd2012FCIQMCjellium}.

Calculations can be made more challenging by considering the system at non-zero temperature, which acts as a model for the interiors of stars and planets, or for laser-ignited plasma used in fusion experiments~\cite{dornheim2017jelliumFiniteTempReview, dornheim2018uniform}. Alternatively, we can consider additional interactions, such as spin-orbit coupling~\cite{liu2020PhaseDiagramHEGwithSpinOrbit}. \\

The uniform electron gas has previously been identified as a candidate system for quantum phase estimation~\cite{babbush2018planewaves} due to the desire to seek accurate, bias-free ground state energies. Existing resource estimates for applying phase estimation to Jellium~\cite{babbush2018EncodingElectronicSpectra, Kivlichan2020improvedfault} project the Jellium Hamiltonian onto the plane wave dual basis, and so can be directly compared with FCIQMC methods. As discussed above, these calculations must first be extrapolated to the basis set limit, before extrapolation to the thermodynamic limit is performed. Previous estimates~\cite{babbush2018EncodingElectronicSpectra,Kivlichan2020improvedfault} have only considered the quantum resources required for phase estimation at half-filling ($\eta = N/2$). However, the most challenging calculation performed in a realistic study will be that with the largest computationally feasible $\eta$ value, subject to the constraint that $N \gg \eta$. Without this constraint, it will not be possible to perform an accurate extrapolation to the continuum limit. In this work, we explicitly consider this regime of interest, and make use of the fermionic seminorm bounds presented in \sec{DecompsAndBounds} to reduce estimates of the Trotter error bound, compared to the state-of-the-art~\cite{Kivlichan2020improvedfault}.

\section{Numerical results}\label{Sec:NumericalResults}

\subsection{Trotter error comparison}\label{Subsec:TrotterErrorComp}

We have numerically evaluated the Trotter error bounds derived in \sec{DecompsAndBounds} for 2D uniform electron gas systems with up to 49 electrons in 512 plane wave dual spin-orbitals. These calculations were performed as outlined in \append{ComputationDetails}, with the help of subroutines present in OpenFermion~\cite{mcclean2020openfermion}, an electronic structure package for quantum computational chemistry\footnote{In Figs. 1 \& 2 we have corrected a small error from the published version of this manuscript, which results in a reduction of the plotted spectral decomposition bounds by a small amount.}. \\

\begin{figure}[!h]
\begin{center}
\includegraphics[width=0.9\columnwidth]{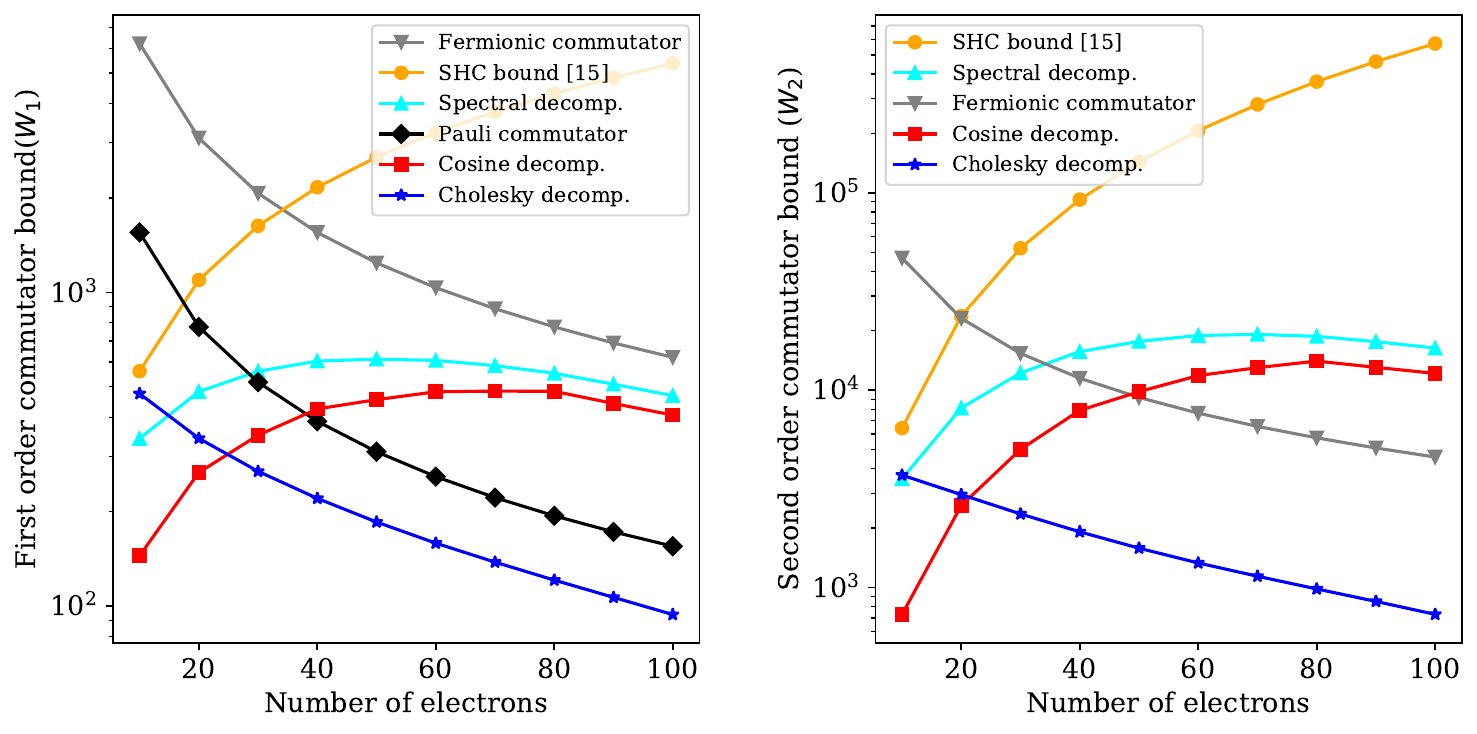}
\caption{First (left) and second (right) order commutator bounds for a 2D uniform electron gas system with $r_s=5$, resolved with 200 spin-orbitals. The electronic density is kept fixed, such that the volume of the simulation cell increases with the number of electrons considered, which alters the Hamiltonian coefficients. This effect competes with the electron number-dependence of the fermionic seminorm to determine the resulting error bounds. For the Cholesky decomposition the chemical potential was shifted by the minimum value that ensured $V$ was positive definite.}
\label{Fig:FixedOrbs}
\end{center}
\end{figure}

In \fig{FixedOrbs} we plot the first ($W_1$) and second ($W_2$) order commutator bounds for the Hamiltonian decompositions discussed in this work. We consider a simulation cell resolved with 200 spin-orbitals, and vary the number of electrons in the cell. The Wigner-Seitz radius is set to $r_s=5$. We fix the electron density, such that the volume of the simulation cell increases proportionally with the number of electrons considered. As the cell volume increases, the Hamiltonian coefficients decrease in magnitude. This effect will contribute to a reduction of the commutator bound. However, increasing the number of electrons in the system also increases the number of eigenvalues considered when taking the fermionic seminorm of the relevant free-fermionic coefficient matrices. This effect increases the commutator bound. The competition between these effects can lead to non-trivial behaviour as the number of electrons is varied -- this is particularly evident for the spectral and cosine decomposition bounds. These are the decompositions that maximally exploit the fermionic seminorm, leading to their improved behaviour in the low-filling regime. In contrast, the Pauli and Fermionic commutator bounds receive no benefit from decreasing fermion number $\eta$. However, we see that these bounds, as well as our Cholesky bound (which only makes partial use of the fermionic seminorm) perform well close to half-filling, due to their sensitive dependence on the Hamiltonian coefficients. Although it is masked by the log-scale used in the plots, we observe that for a fixed number of spin-orbitals $||T|| \propto 1/\eta$ and $||V||_{\mathrm{max}} \in \mathcal{O}(1)$, so the first-order SHC bound is proportional to $\eta$, and the second-order SHC bound is proportional to $\eta^2$. While the SHC bound exploits the fermionic seminorm, it does not fully exploit the reduction in Hamiltonian coefficient magnitudes at high-filling fractions. \\

\begin{figure}[!h]
\begin{center}
\includegraphics[width=1.0\columnwidth]{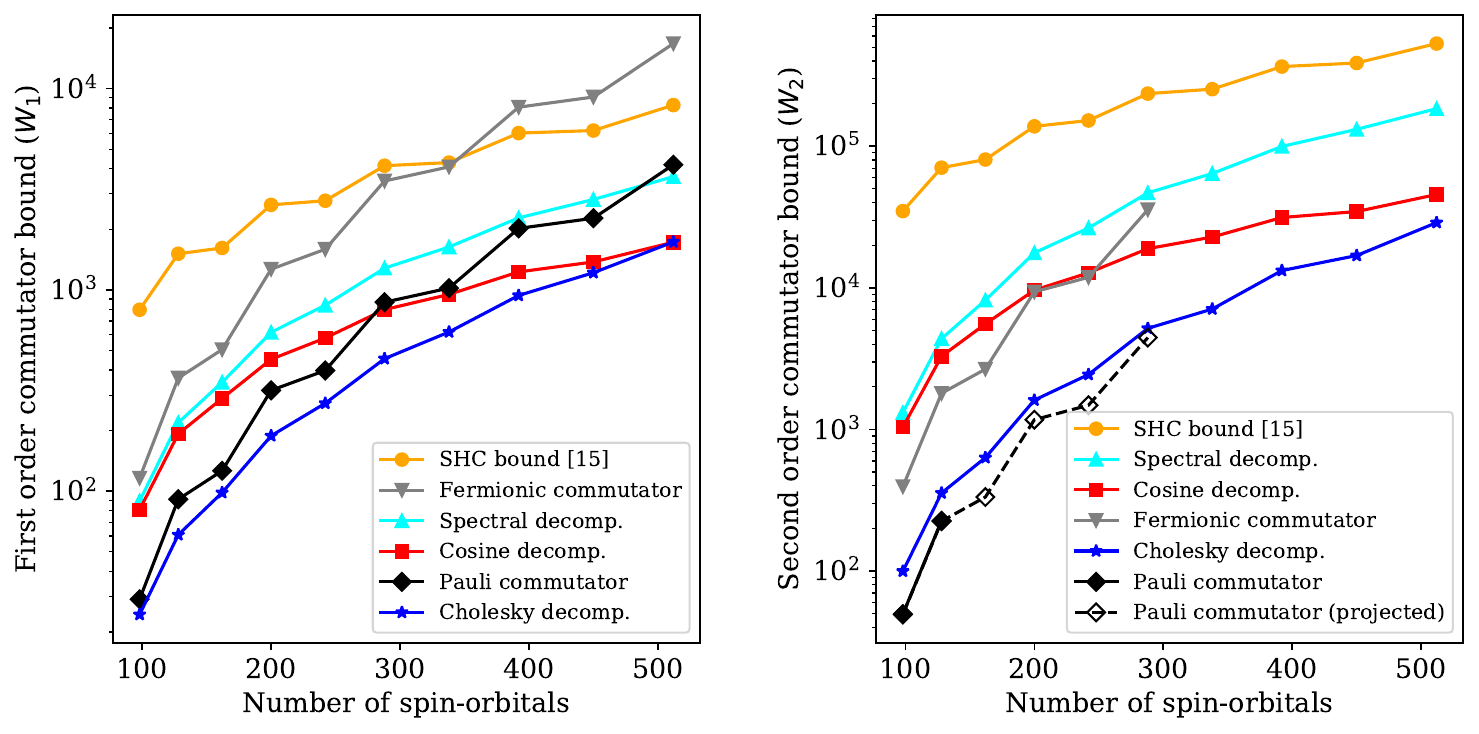}
\caption{First (left) and second (right) order commutator bounds for a 2D uniform electron gas system with $r_s=5$, and 49 electrons, as a function of the number of spin-orbitals used to resolve the system. Fermionic and Pauli commutator bounds could not be calculated for all datapoints, due to the large memory requirements of those approaches. The `projected Pauli bounds' were obtained as described in \append{ProjectedPauli}. For the Cholesky decomposition the chemical potential was shifted by the minimum value that ensured $V$ was positive definite.}
\label{Fig:fixed_electrons}
\end{center}
\end{figure}

In \fig{fixed_electrons} we again plot the first and second-order commutator bounds, but here keep the fermion number fixed at $\eta=49$ (as well as fixing $r_s=5$, and the cell volume) and instead vary the number of spin-orbitals used. The second-order Pauli and Fermionic commutator bounds were only calculated up to 128 and 288 spin-orbitals, respectively, as the memory required for these calculations was prohibitive beyond this point. We have extrapolated the performance of the second-order Pauli bounds to larger $N$ values, as described in \append{ProjectedPauli}. Performing simulations with a fixed number of electrons, while increasing the number of spin-orbitals, would enable us to perform extrapolation to the basis-set limit. We observe that close to half-filling, the Cholesky and Pauli bounds outperform all others considered. However, the cosine decomposition performs best in the low-filling fraction regime, due to its increased exploitation of the fermionic seminorm.

\subsection{Phase estimation resource estimates}\label{Subsec:PhaseEstimation}

In this section, we discuss the resources required for performing Trotter-based phase estimation on the uniform electron gas systems discussed in the previous section. Our cost estimates focus on the number of logical qubits and $T$ \& Toffoli gates required (as these are the dominant factors in surface code-based resource estimates), and neglect the costs of Clifford gates. Our approach closely follows that of Ref.~\cite{Kivlichan2020improvedfault}, with an improved use of Hamming weight phasing (HWP)~\cite{campbell2020early}.

We distribute the total budget for error in energy estimation ($\delta=\Delta_{PE}+\Delta_{TS}+\Delta_{\mathrm{syn}}$) roughly as follows: 33\% to Trotter error $\Delta_{TS}$; 66\% to phase estimation error $\Delta_{PE}$; and 1\%  to rotation synthesis error $\Delta_{\mathrm{syn}}$. In practice, we numerically optimise the error budget allocated to rotation synthesis error, but the optimal choice only differs slightly from 1\%.  With this split of the error budget, one finds~\cite{Kivlichan2020improvedfault,campbell2020early} that we need $N_{PE}=\tilde{\mathcal{O}}( W_2^{1/2}/\delta^{3/2})$ (the tilde in our $\tilde{\mathcal{O}}$ notation denotes that logarithmic factors have been suppressed and also hides constant factors) Trotter steps. As we outline in \append{PhaseEstErrorBudget}, each Trotter step can be implemented with $\tilde{\mathcal{O}}(N^2)$ non-Clifford gates for an $N$ spin-orbital problem. Therefore, the total algorithm complexity is $\tilde{\mathcal{O}}(N^2 W_2^{1/2} /\delta^{3/2}) $ where $W_2$ contains some dependence on $N$ and $\eta$. The primary focus of our work has been to tighten the values of $W_2$ and we expect a factor $C$ reduction in $W_2$ will lead to a corresponding factor $C^{1/2}$ runtime improvement.\\

We numerically count the non-Clifford resources for a range of different $r_s$, $\eta$, and $N$ values, as shown in \tab{JelliumPhaseEstDataComparison}. We consider an architecture that distills $T$ gates as its non-Clifford resource. We compare the gate counts obtained by our Trotter-based approach to those obtained using the Trotter-based approach of Ref.~\cite{Kivlichan2020improvedfault} (which used the fermionic commutator bound on the Trotter error) and those obtained using the qubitization-based method of Ref.~\cite{babbush2018EncodingElectronicSpectra}. We consider an extensive error bound $\delta = 1$~mHartree per electron, consistent with leading classical approaches~\cite{ruggeri2018FCIQMCspinpolarisedJellium, shepherd2012investigation,shepherd2012FCIQMCjellium}. It is too memory intensive to calculate the fermionic commutator (`FC') bounds for $16 \times 16$ systems, showing the limitations of the prior art.

\begin{table*}[]
    \centering
   \begin{tabular}{cccc|c|c|cc}
 &  & Filling fraction & Size &  Our best  & FC & \multicolumn{2}{c}{Qubitization} \\ 
   $r_s$ & $\eta$  & $\eta / 2 L_X L_ Y  $ & $L_X \times L_Y$  & $N_T+4 N_{\mathrm{tof}}$  & $N_T+4 N_{\mathrm{tof}}$   & $N_T+4 N_{\mathrm{tof}}$ & Anc.  \\ \hline
   5 & 49 & 0.10 & 16 $\times$ 16 & $2.2 \times 10^{11}$ &  No data  & $2.0 \times 10^{10}$ & 105  \\
   5 & 49 & 0.17 & 12$\times$12 & $3.2 \times 10^{10}$ & $8.8 \times 10^{10}$ & $3.5 \times 10^{9}$ & 96 \\
   5 & 49 & 0.19 & 16$\times$8 &  $1.7 \times 10^{10}$ & $4.8 \times 10^{10}$ & $2.5 \times 10^9$ & 94 \\
   5 & 49 & 0.38 & 8$\times$8 &  $1.3 \times 10^9$ & $3.0 \times 10^9$ & $3.2 \times 10^{8}$ & 83 \\  \hline
   10 & 10 & 0.02 & 16$\times$16 & $3.4 \times 10^{11}$ & No data & $9.6 \times 10^{10}$ & 112 \\
   10 & 10 & 0.03 & 12$\times$12 &  $8.6 \times 10^{10}$ & $1.1 \times 10^{12}$ & $1.7 \times 10^{10}$ & 103  \\
   10 & 10 & 0.04 & 16$\times$8 & $5.4 \times 10^{10}$ & $6.2 \times 10^{11}$ & $1.2 \times 10^{10}$ & 101 \\
   10 & 10 & 0.08 & 8$\times$8 &  $8.1 \times 10^9$ & $3.9 \times 10^{10}$ &  $1.6 \times 10^{9}$ & 90 \\
   10 & 49 & 0.10 & 16$\times$16 & $1.1 \times 10^{11}$ & No data & $2.0 \times 10^{10}$ & 105\\
   10 & 49 & 0.17 & 12$\times$12 & $1.6 \times 10^{10}$ & $4.3 \times 10^{10}$ & $3.5 \times 10^{9}$ & 96 \\
   10 & 49 & 0.19 & 16$\times$8 &  $8.6 \times 10^9$ & $2.4 \times 10^{10}$ & $2.5 \times 10^9$ & 94 \\
   10 & 49 & 0.38 & 8$\times$8 &  $6.5 \times 10^8$ & $1.5 \times 10^{9}$ & $3.2 \times 10^{8}$ & 83 \\ \hline
  \end{tabular}
    \caption{A comparison of resource estimates for phase estimation of Jellium, using three different methods. We consider an energy error budget of $\delta=1$ mHa per electron. `Our best' refers to the Trotter-based phase estimation considered in this work, using our best bound for the Trotter error. `FC' refers to the Trotter-based phase estimation considered in Ref.~\cite{Kivlichan2020improvedfault}, which uses the fermionic commutator bound for the Trotter error. The fermionic commutator (FC) bounds are too memory intensive to be calculated for the $16 \times 16$ systems. In both Trotter methods, 16 additional qubits are used (14 for Hamming weight phasing, one for phase estimation, and one for gate synthesis). `Qubitization' refers to the post-Trotter approach considered in Ref.~\cite{babbush2018EncodingElectronicSpectra} (which we discuss in \append{Qubitization}). Four $T$ gates can be used to implement a Toffoli gate, so the total aggregated $T$ count for the algorithm is  $N_{T}+4N_{\mathrm{tof}}$.}
    \label{tab:JelliumPhaseEstDataComparison}
\end{table*}

Comparing the gate counts obtained using our novel Trotter error bounds to those obtained using the existing fermionic commutator bound, we observe a reduction in $T$ count by a factor of between $2.3 - 12.7\times$. This improvement is more pronounced at lower filling fractions, demonstrating the anticipated benefit of using the fermionic seminorm. The largest of these improvements stems from a reduction in Trotter error by a factor of 150 for $\eta=10,~N=288,~r_s=10$. In the high accuracy regime of 49 electrons in $\sim 1000$ spin-orbitals, we would expect our bounds to provide an order-of-magnitude improvement over the prior art as this would be similar to the improvements showcased by our $\eta=10$, $N=288$ results.

Comparing our results to those of qubitization, we see that qubitization consistently (for $r_s =5, 10$) achieves a lower $T$ count for the systems considered, by a factor of $2
- 11\times$. This comes at a cost of using $5 - 7\times$ more ancilla qubits. We show in \append{Qubitization} that for 2D Jellium at large $r_s$ values, the cost of qubitization is roughly independent of $r_s$ (when $N, \eta$ are fixed). In contrast, the cost of our Trotter-based approach scales as $1/r_s$. These scalings are evident in Table.~\ref{tab:JelliumPhaseEstDataComparison}. As such, the Trotter-based approach will be the more suitable method for calculations probing the phase diagram of 2D Jellium, which target $r_s \geq 20$~\cite{drummond2009PhaseDiagram2DHEG}. In contrast, qubitization will likely perform better for the warm, dense phase ($r_s < 1$~\cite{dornheim2017jelliumFiniteTempReview,dornheim2018uniform}). Our Trotter-based approach also scales less efficiently with target error than qubitization ($\delta^{-3/2}$ vs $\delta^{-1}$), and so the advantage of qubitization will also decrease if the target error in our calculations is loosened.\\

As a final caveat, this analysis assumes that we can prepare the main register in the desired energy eigenstate. If we are only able to prepare a state with overlap $\gamma < 1$, then the circuit depth required is increased by a factor of $1/\gamma$. For the sake of comparison with prior art, we assume that $\gamma=1$, but note that it is an open question whether an eigenstate with sufficient overlap can be prepared~\cite{mcclean2014locality,babbush2015chemical,tubman2018postponing}. It will be necessary to repeat the phase estimation process a number of times, to ascertain that phase estimation has found the desired eigenstate. One can also consider other methods of phase estimation, such as that of Ref.~\cite{lin2021heisenberg}, which requires an increased number of repetitions of the algorithm, but that has coherent circuit depth independent of $\gamma$.

\section{Discussion}\label{Sec:Discussion}
We have demonstrated a substantial benefit of our approach to calculating Trotter errors, both in terms of tightness of the bound and the classical runtime and memory complexity.  We have primarily focused on second-order Trotter in the plane wave dual basis, but our techniques naturally generalize. For more compact basis sets, fewer orbitals are required, but the Hamiltonian contains $\mathcal{O}(N^4)$ terms instead of $\mathcal{O}(N^2)$. In such a compact basis set, the spectral and Cholesky decompositions are still applicable~\cite{motta2018low}, but it is unclear whether an analogue of the cosine decomposition could be used to obtain an even tighter bound in the low-filling fraction regime. Fourth-order Trotter may produce results competitive with those here~\cite{Childs2018towards,childs2019trottererror}, if $5^4 \cdot W_4 < W_2^2 $, and a similarly low-overhead compilation of the Trotter circuit can be found. While the methods introduced in this work apply straightforwardly to higher-order Trotter, calculating the fourth-order bounds would require time scaling as $\mathcal{O}(N^7)$, making it a potentially costly endeavour.

While this work has focused on the performance of Trotter methods, so-called post-Trotter methods~\cite{childs2012hamiltonian,Berry15,babbush2016exponential,QSP17,babbush2018EncodingElectronicSpectra,meister2020tailoring} are known to have superior asymptotic performance with respect to target error. These methods have also leveraged Hamiltonian factorizations to reduce costs~\cite{Berry2019qubitization,vonburg2020catalysis,lee2020TensorHyperContraction}. Trotter methods often possess good constant prefactors in the runtime and require few additional ancilla qubits, compared to post-Trotter methods. As such,
it has been proposed~\cite{Kivlichan2020improvedfault,campbell2020early} that Trotter methods could perform better at some tasks in the pre-asymptotic regime. The gate counts presented in Sec.~\ref{Subsec:PhaseEstimation} show that our Trotter approach can be competitive with post-Trotter methods like qubitization, in some regimes of interest, and will even use fewer gates than qubitization for large enough Wigner-Seitz radius.  It is currently unclear whether second quantized post-Trotter methods can similarly exploit low-filling fractions, which appears to strengthen the case for Trotter methods in this regime. Working in first quantization, one could certainly exploit low-filling fractions, but quantum algorithms would need to be substantially modified to work in this setting~\cite{babbush2019sublinear, su2021fault}.

\section{Acknowledgements}

We thank Hsin-Yuan (Robert) Huang, Fernando Brandao, Mario Berta and Michael Kastoryano for discussions through this project.  Yuan Su's contribution to this project was made while at Caltech. He was supported in part by the National Science Foundation RAISE-TAQS 1839204 and Amazon Web Services, AWS Quantum Program. The Institute for Quantum Information and Matter is an NSF Physics Frontiers Center PHY-1733907.

\bibliographystyle{unsrtnat}
\bibliography{Bibliography}

\appendix

\section{SHC bounds}\label{App:AnalyticBounds}

We consider simulating the following class of interacting electrons
\begin{equation}
\label{eq:second_quantized_ham}
H=H_t+H_v:=\sum_{j,k}T_{j,k}a_j^\dagger a_k+\sum_{l,m}V_{l,m}n_l n_m,
\end{equation}
where $a_j^\dagger$ and $a_k$ are the fermionic \emph{creation} and \emph{annihilation operators}, $n_l$ are the \emph{occupation-number operators}, $T$ and $V$ are coefficient matrices, and the summation is over $N$ spin orbitals. We seek to bound the fermionic seminorm of the nested commutators $\left[H_t,\left[H_t, H_v\right]\right]$ and $\left[H_v,\left[H_v, H_t\right]\right]$. 

We know from [Ref.~\cite{su2020nearly}, Eq.\ (60)] that
\begin{equation}
\begin{aligned}
\left[H_t, H_v\right]
&=\sum_{j,k,m}T_{j,k}V_{k,m}a_j^\dagger n_ma_k
+\sum_{j,k}T_{j,k}V_{k,k}a_j^\dagger a_k
+\sum_{j,k,l}T_{j,k}V_{l,k}a_j^\dagger n_la_k\\
&\quad -\sum_{j,k,m}T_{j,k}V_{j,m}a_j^\dagger n_ma_k
-\sum_{j,k}T_{j,k}V_{j,j}a_j^\dagger a_k
-\sum_{j,k,l}T_{j,k}V_{l,j}a_j^\dagger n_l a_k.
\end{aligned}
\end{equation}
Applying [Ref.~\cite{su2020nearly}, Eq.\ (77)], we get the following expansion
\begin{align}
	&\left[H_t,\left[H_t, H_v\right]\right]\\
	=&\sum_{j,k,m}T_{j,k}V_{k,m}\left(\sum_{j',k'}\delta_{k',j}T_{j',k'}a_{j'}^\dagger \right) n_ma_k
	+\sum_{j,k}T_{j,k}V_{k,k}\left(\sum_{j',k'}\delta_{k',j}T_{j',k'}a_{j'}^\dagger \right) a_k \nonumber \\
	&+ \sum_{j,k,l}T_{j,k}V_{l,k}\left(\sum_{j',k'}\delta_{k',j}T_{j',k'}a_{j'}^\dagger \right) n_la_k
	-\sum_{j,k,m}T_{j,k}V_{j,m}\left(\sum_{j',k'}\delta_{k',j}T_{j',k'}a_{j'}^\dagger \right) n_ma_k \nonumber\\
	&-\sum_{j,k}T_{j,k}V_{j,j}\left(\sum_{j',k'}\delta_{k',j}T_{j',k'}a_{j'}^\dagger \right) a_k
	-\sum_{j,k,l}T_{j,k}V_{l,j}\left(\sum_{j',k'}\delta_{k',j}T_{j',k'}a_{j'}^\dagger \right) n_l a_k \nonumber \\
	&\ -\sum_{j,k,m}T_{j,k}V_{k,m}a_j^\dagger n_m\left(\sum_{j',k'}T_{j',k'}\delta_{j',k}a_{k'}\right)
	-\sum_{j,k}T_{j,k}V_{k,k}a_j^\dagger \left(\sum_{j',k'}T_{j',k'}\delta_{j',k}a_{k'}\right) \nonumber \\
	&-\sum_{j,k,l}T_{j,k}V_{l,k}a_j^\dagger n_l\left(\sum_{j',k'}T_{j',k'}\delta_{j',k}a_{k'}\right)
	+\sum_{j,k,m}T_{j,k}V_{j,m}a_j^\dagger n_m\left(\sum_{j',k'}T_{j',k'}\delta_{j',k}a_{k'}\right) \nonumber \\
	&+\sum_{j,k}T_{j,k}V_{j,j}a_j^\dagger \left(\sum_{j',k'}T_{j',k'}\delta_{j',k}a_{k'}\right)
	+\sum_{j,k,l}T_{j,k}V_{l,j}a_j^\dagger n_l \left(\sum_{j',k'}T_{j',k'}\delta_{j',k}a_{k'}\right) \nonumber \\
	&\ +\sum_{j,k,m}T_{j,k}V_{k,m}a_j^\dagger \left(\sum_{j',k'}T_{j',k'}\delta_{k',m}a_{j'}^\dagger a_{k'}\right)a_k
	+\sum_{j,k,l}T_{j,k}V_{l,k}a_j^\dagger \left(\sum_{j',k'}T_{j',k'}\delta_{k',l}a_{j'}^\dagger a_{k'}\right)a_k \nonumber \\
	&\ -\sum_{j,k,m}T_{j,k}V_{j,m}a_j^\dagger \left(\sum_{j',k'}T_{j',k'}\delta_{k',m}a_{j'}^\dagger a_{k'}\right)a_k
	-\sum_{j,k,l}T_{j,k}V_{l,j}a_j^\dagger \left(\sum_{j',k'}T_{j',k'}\delta_{k',l}a_{j'}^\dagger a_{k'}\right) a_k \nonumber \\
	&\ -\sum_{j,k,m}T_{j,k}V_{k,m}a_j^\dagger \left(\sum_{j',k'}T_{j',k'}\delta_{j',m}a_{j'}^\dagger a_{k'}\right)a_k
	-\sum_{j,k,l}T_{j,k}V_{l,k}a_j^\dagger \left(\sum_{j',k'}T_{j',k'}\delta_{j',l}a_{j'}^\dagger a_{k'}\right)a_k \nonumber \\
	&\ +\sum_{j,k,m}T_{j,k}V_{j,m}a_j^\dagger \left(\sum_{j',k'}T_{j',k'}\delta_{j',m}a_{j'}^\dagger a_{k'}\right)a_k
	+\sum_{j,k,l}T_{j,k}V_{l,j}a_j^\dagger \left(\sum_{j',k'}T_{j',k'}\delta_{j',l}a_{j'}^\dagger a_{k'}\right) a_k,
\end{align}
which implies through [Ref.~\cite{su2020nearly}, Proposition 10]
\begin{equation}
\begin{aligned}
	\norm{\left[H_t,\left[H_t,H_v\right]\right]}_\eta
	&\leq\norm{T}^2\eta\norm{V}_{\max}\eta+\norm{T}^2\eta\norm{V}_{\max}+\norm{T}^2\eta\norm{V}_{\max}\eta\\
	&\quad+\norm{T}^2\eta\norm{V}_{\max}\eta+\norm{T}^2\eta\norm{V}_{\max}+\norm{T}^2\eta\norm{V}_{\max}\eta\\
	&\quad+\norm{T}^2\eta\norm{V}_{\max}\eta+\norm{T}^2\eta\norm{V}_{\max}+\norm{T}^2\eta\norm{V}_{\max}\eta\\
	&\quad+\norm{T}^2\eta\norm{V}_{\max}\eta+\norm{T}^2\eta\norm{V}_{\max}+\norm{T}^2\eta\norm{V}_{\max}\eta\\
	&\quad+\left(\norm{T}\eta\right)^2\norm{V}_{\max}+\left(\norm{T}\eta\right)^2\norm{V}_{\max}\\
	&\quad+\left(\norm{T}\eta\right)^2\norm{V}_{\max}+\left(\norm{T}\eta\right)^2\norm{V}_{\max}\\
	&\quad+\left(\norm{T}\eta\right)^2\norm{V}_{\max}+\left(\norm{T}\eta\right)^2\norm{V}_{\max}\\
	&\quad+\left(\norm{T}\eta\right)^2\norm{V}_{\max}+\left(\norm{T}\eta\right)^2\norm{V}_{\max}\\
	&\leq16\norm{T}^2\norm{V}_{\max}\eta^2+4\norm{T}^2\norm{V}_{\max}\eta.
\end{aligned}
\end{equation}
Similarly, we have from [Ref.~\cite{su2020nearly}, Eq.\ (78)]
\begin{align}
	&\left[H_v,\left[H_t, H_v\right]\right]  \\
	=&\sum_{j,k,m}T_{j,k}V_{k,m}a_j^\dagger \left(\sum_{l',m'}V_{l',m'}\delta_{m',j}n_{l'}\right)n_ma_k
	+\sum_{j,k}T_{j,k}V_{k,k}a_j^\dagger \left(\sum_{l',m'}V_{l',m'}\delta_{m',j}n_{l'}\right)a_k \notag \\
	&+\sum_{j,k,l}T_{j,k}V_{l,k}a_j^\dagger \left(\sum_{l',m'}V_{l',m'}\delta_{m',j}n_{l'}\right)n_la_k
	 -\sum_{j,k,m}T_{j,k}V_{j,m}a_j^\dagger \left(\sum_{l',m'}V_{l',m'}\delta_{m',j}n_{l'}\right)n_ma_k \notag \\
	&-\sum_{j,k}T_{j,k}V_{j,j}a_j^\dagger \left(\sum_{l',m'}V_{l',m'}\delta_{m',j}n_{l'}\right)a_k
	-\sum_{j,k,l}T_{j,k}V_{l,j}a_j^\dagger \left(\sum_{l',m'}V_{l',m'}\delta_{m',j}n_{l'}\right)n_l a_k \notag\\
	&\ +\sum_{j,k,m}T_{j,k}V_{k,m}a_j^\dagger \left(\sum_{l',m'}V_{l',m'}\delta_{l',j}n_{m'}\right)n_ma_k
	+\sum_{j,k}T_{j,k}V_{k,k}a_j^\dagger \left(\sum_{l',m'}V_{l',m'}\delta_{l',j}n_{m'}\right)a_k \notag\\
	&+\sum_{j,k,l}T_{j,k}V_{l,k}a_j^\dagger \left(\sum_{l',m'}V_{l',m'}\delta_{l',j}n_{m'}\right)n_la_k
	-\sum_{j,k,m}T_{j,k}V_{j,m}a_j^\dagger \left(\sum_{l',m'}V_{l',m'}\delta_{l',j}n_{m'}\right)n_ma_k \notag\\
	&-\sum_{j,k}T_{j,k}V_{j,j}a_j^\dagger \left(\sum_{l',m'}V_{l',m'}\delta_{l',j}n_{m'}\right)a_k
	-\sum_{j,k,l}T_{j,k}V_{l,j}a_j^\dagger \left(\sum_{l',m'}V_{l',m'}\delta_{l',j}n_{m'}\right)n_l a_k \notag\\
	&\ +\sum_{j,k,m}T_{j,k}V_{k,m}a_j^\dagger \left(\sum_{l',m'}V_{l',m'}\delta_{m',j}\delta_{l',j}\right)n_ma_k
	+\sum_{j,k}T_{j,k}V_{k,k}a_j^\dagger \left(\sum_{l',m'}V_{l',m'}\delta_{m',j}\delta_{l',j}\right)a_k \notag\\
	&+\sum_{j,k,l}T_{j,k}V_{l,k}a_j^\dagger \left(\sum_{l',m'}V_{l',m'}\delta_{m',j}\delta_{l',j}\right)n_la_k 
	-\sum_{j,k,m}T_{j,k}V_{j,m}a_j^\dagger \left(\sum_{l',m'}V_{l',m'}\delta_{m',j}\delta_{l',j}\right)n_ma_k \notag\\
	&-\sum_{j,k}T_{j,k}V_{j,j}a_j^\dagger \left(\sum_{l',m'}V_{l',m'}\delta_{m',j}\delta_{l',j}\right)a_k
	-\sum_{j,k,l}T_{j,k}V_{l,j}a_j^\dagger \left(\sum_{l',m'}V_{l',m'}\delta_{m',j}\delta_{l',j}\right)n_l a_k \notag\\
	&\ -\sum_{j,k,m}T_{j,k}V_{k,m}a_j^\dagger n_m\left(\sum_{l',m'}V_{l',m'}\delta_{m',k}n_{l'}\right)a_k
	-\sum_{j,k}T_{j,k}V_{k,k}a_j^\dagger \left(\sum_{l',m'}V_{l',m'}\delta_{m',k}n_{l'}\right)a_k \notag\\
	& -\sum_{j,k,l}T_{j,k}V_{l,k}a_j^\dagger n_l\left(\sum_{l',m'}V_{l',m'}\delta_{m',k}n_{l'}\right)a_k
	 +\sum_{j,k,m}T_{j,k}V_{j,m}a_j^\dagger n_m\left(\sum_{l',m'}V_{l',m'}\delta_{m',k}n_{l'}\right)a_k \notag\\
	& +\sum_{j,k}T_{j,k}V_{j,j}a_j^\dagger \left(\sum_{l',m'}V_{l',m'}\delta_{m',k}n_{l'}\right)a_k
	+\sum_{j,k,l}T_{j,k}V_{l,j}a_j^\dagger n_l \left(\sum_{l',m'}V_{l',m'}\delta_{m',k}n_{l'}\right)a_k \notag\\
	&\ -\sum_{j,k,m}T_{j,k}V_{k,m}a_j^\dagger n_m\left(\sum_{l',m'}V_{l',m'}\delta_{l',k}n_{m'}\right)a_k
	-\sum_{j,k}T_{j,k}V_{k,k}a_j^\dagger \left(\sum_{l',m'}V_{l',m'}\delta_{l',k}n_{m'}\right)a_k \notag\\
	& -\sum_{j,k,l}T_{j,k}V_{l,k}a_j^\dagger n_l\left(\sum_{l',m'}V_{l',m'}\delta_{l',k}n_{m'}\right)a_k
    +\sum_{j,k,m}T_{j,k}V_{j,m}a_j^\dagger n_m\left(\sum_{l',m'}V_{l',m'}\delta_{l',k}n_{m'}\right)a_k \notag\\
	&+\sum_{j,k}T_{j,k}V_{j,j}a_j^\dagger \left(\sum_{l',m'}V_{l',m'}\delta_{l',k}n_{m'}\right)a_k
	+\sum_{j,k,l}T_{j,k}V_{l,j}a_j^\dagger n_l \left(\sum_{l',m'}V_{l',m'}\delta_{l',k}n_{m'}\right)a_k \notag\\
	&\ -\sum_{j,k,m}T_{j,k}V_{k,m}a_j^\dagger n_m\left(\sum_{l',m'}V_{l',m'}\delta_{m',k}\delta_{l',k}\right)a_k
	-\sum_{j,k}T_{j,k}V_{k,k}a_j^\dagger \left(\sum_{l',m'}V_{l',m'}\delta_{m',k}\delta_{l',k}\right)a_k \notag\\
	&-\sum_{j,k,l}T_{j,k}V_{l,k}a_j^\dagger n_l\left(\sum_{l',m'}V_{l',m'}\delta_{m',k}\delta_{l',k}\right)a_k
	+\sum_{j,k,m}T_{j,k}V_{j,m}a_j^\dagger n_m\left(\sum_{l',m'}V_{l',m'}\delta_{m',k}\delta_{l',k}\right)a_k \notag\\
	&+\sum_{j,k}T_{j,k}V_{j,j}a_j^\dagger \left(\sum_{l',m'}V_{l',m'}\delta_{m',k}\delta_{l',k}\right)a_k
	+\sum_{j,k,l}T_{j,k}V_{l,j}a_j^\dagger n_l \left(\sum_{l',m'}V_{l',m'}\delta_{m',k}\delta_{l',k}\right)a_k,
\end{align}
which implies
\begin{equation}
	\norm{\left[H_v,\left[H_t, H_v\right]\right]}_\eta
	\leq24\norm{T}\norm{V}_{\max}^2\eta^3+12\norm{T}\norm{V}_{\max}^2\eta^2.
\end{equation}

\section{Spectral decompositions}
\label{App:Spectral}

Here we review how a general electronic structure Hamiltonian can be factorised using spectral decompositions, with slight modifications allowing Cholesky decompositions to be used.  If the Hamiltonian is given in the form
\begin{align}
    H & = \sum_{pqrs} h_{p q r s} a^\dagger_p a^\dagger_q a_r a_s ,
\end{align}
then we first use the fermionic anti-commutation rules to rewrite it in ``chemist notation" as follows
\begin{align}
    H & = - \sum_{pqrs} h_{p q r s} a^\dagger_p a^\dagger_q a_s a_r  , \\
    & = - \sum_{pqrs} h_{p q r s} a^\dagger_p (\delta_{q,s} -  a_s  a^\dagger_q) a_r
\end{align}
We split $H$ into $H = H' - H_0$, where
\begin{align}
    H' & = \sum_{pqrs} h_{p q r s} a^\dagger_p   a_s  a^\dagger_q a_r = 
    \sum_{pqrs} h_{p r s q} a^\dagger_p   a_q  a^\dagger_r a_s
    = \sum_{pqrs} V_{p q r s} a^\dagger_p a_q a^\dagger_r   a_s \\
    H_0 & =  \sum_{pqrs} h_{p q r s} a^\dagger_p  a_r  \delta_{q,s} ,
\end{align}
where in $H'$ we have changed variables so that $s \rightarrow q \rightarrow r \rightarrow s$ and introduced $V_{p q r s} := h_{p r s q}$. Since $H_0$ is free fermionic, we aim to find a factorization of $H'$. We define a matrix $V_{(pq),(sr)}=V_{pqrs}$ with composite indices $(pq)$ and $(sr)$ so that
\begin{align}
    H' & = \sum_{(pq),(sr)} V_{(pq),(sr)} a^\dagger_p a_q a^\dagger_r   a_s .
\end{align}
Hermiticity of $H'$ entails we can always choose $V_{(pq),(sr)} $ to be Hermitian so that $V_{(pq),(sr)} =V_{(sr),(pq)} ^*$.  Indeed, if $V_{(pq),(sr)} $ is not initially Hermitian, we can always map $V_{(pq),(sr)}  \rightarrow (1/2)(V_{(pq),(sr)}  + V_{(sr),(pq)}^*)$ and confirm that this transformation results in the same Hermitian $H'$.  Therefore, we can diagonalize the matrix with elements $V_{(pq),(sr)}$  so that
\begin{align} \label{Eq:OrigDecomp}
    V_{pqrs}=V_{(pq),(sr)} & = \sum_{\ell}  u_{pq,\ell} \lambda_{\ell} u_{sr, \ell}^{*} ,
\end{align}
where $\lambda_{\ell}$ are real eigenvalues and $u_{i,j}$ are matrix elements of a unitary $U$. Substituting this into the expressions for $H'$ we get
\begin{align} \label{Eq:HprimeExpand}
    H' & = \sum_{\ell=1}^L \lambda_{\ell} \left(  \sum_{pq} u_{pq,\ell} a^\dagger_p a_q \right) \left(\sum_{sr} u_{sr,\ell}^*  a^\dagger_r   a_s \right)
\end{align}
We define $\mathcal{L}\defindex{\ell} := \sum_{pq} u_{pq,\ell}  a^\dagger_p   a_q$ which is the first bracketed factor above. Notice that by changing dummy variables in the summation $p \rightarrow s$ and $q \rightarrow r$ we also have $\mathcal{L}\defindex{\ell}= \sum_{sr} u_{sr,\ell}  a^\dagger_s   a_r $.  Taking the Hermitian conjugate, we have $\mathcal{L}\defindex{\ell}^\dagger= \sum_{sr} u_{sr,\ell}^*  a^\dagger_r   a_s $, which corresponds to the second bracketed factor in \Eq{HprimeExpand}. Therefore,
\begin{align}
    H' & = \sum_{\ell=1}^L \lambda_{\ell} \mathcal{L}_{\ell} \mathcal{L}_{\ell}^\dagger
\end{align}
where $\mathcal{L}_{\ell}$ is free-fermionic with coefficient matrix $X_{\ell}$ with matrix elements $[X_{\ell}]_{p,q}=u_{pq,\ell}$. Therefore, $\mathcal{L}_{\ell}= H(X_{\ell} )$ and $\mathcal{L}_{\ell}^\dagger= H(X_{\ell}^\dagger )$.  Therefore,
\begin{align} \label{Eq:FullDecomp}
    H & = -H_0 + \sum_{\ell=1}^L \lambda_{\ell}  H(X_{\ell}) H(X_{\ell}^\dagger).
\end{align}
Computing the fermionic seminorm is significantly easier for $H(A)$ with Hermitian $A$. In general, the individual factors $H(X_{\ell})$ and $H(X_{\ell}^\dagger)$ might not be Hermitian, even though the full \Eq{FullDecomp} is Hermitian.  There are two possible solutions to enforce Hermiticity of the factors.

Following Sec IV. A of \cite{poulin2014trotter}, we can always decompose $X_{\ell}$ in terms of a Hermitian and skew-Hermitian part $X_{\ell}= A_\ell + i B_{\ell}$ so that $A_\ell$ and $B_{\ell}$ are Hermitian. Then for each $\ell$ term we have 
\begin{align} \label{EllTerm}
    H(X_{\ell}) H(X_{\ell}^\dagger) & = (H(A_{\ell})+i H(B_{\ell}))(H(A_{\ell})- i H(B_{\ell})) \\ \nonumber
   & =  H(A_{\ell})^2 -  i [ H(A_{\ell}), H(B_{\ell})] + H(B_{\ell})^2  \\  
   & =  H(A_{\ell})^2 -    H(  i[A_{\ell},B_{\ell} ]) + H(B_{\ell})^2  .
\end{align}
The term $-H(  i[A_{\ell},B_{\ell} ])$ is free-fermionic and Hermitian and so can be added to the free-fermionic part $H_0$.  The full expression for $H$ therefore will have $2L$ terms of the form $H(A_{\ell})^2$ or $H(B_{\ell})^2$.

The above approach is fully general, but results in a doubling of the number of terms in the summation.  In some cases, we can directly ensure Hermiticity of $H(X_{\ell})$ without any increase in the number of terms.  Here we expand on the discussion given in \cite{Berry2019qubitization,motta2018low} but warn the reader that \cite{motta2018low} contains notational errors. When the basis set used for the fermionic orbitals is real-valued (such as for Gaussian basis sets or the plane wave dual basis), then the Hamiltonian constants $V_{pqrs}$ are real-valued and have an 8-fold symmetry~\cite{motta2018low} so that 
\begin{equation} \label{Eq:EightFold}
    V_{pqrs}=V_{srqp}=V_{pqsr}=V_{qprs}=V_{qpsr}=V_{rsqp}=V_{rspq}=V_{srpq}
\end{equation}
Recall that in the original decomposition for $V_{pqrs}$ we had \Eq{OrigDecomp}.  Since the $V_{(pq),(sr)}$ matrix is real and Hermitian, it is therefore diagonalizable by an orthogonal transformation. Since matrix elements of orthogonal transforms are real, we have $u_{rs,\ell}^*=u_{rs,\ell}$ and so
\begin{equation}
   V_{pqrs} =  \sum_{\ell} \lambda_{\ell}  u_{pq,\ell} u_{sr,\ell} 
\end{equation}
Next, we will show that we can always map $X_{\ell} \rightarrow (X_{\ell} + X_{\ell}^\dagger)/2$ and verify that the new decomposition gives the same total Hamiltonian. The transformation $X_{\ell} \rightarrow (X_{\ell} + X_{\ell}^\dagger)/2$ maps $u_{pq,\ell} \rightarrow (u_{pq,\ell}+ u_{qp,\ell}^* )/2 =(u_{pq,\ell}+ u_{qp,\ell} )/2 $ and so
\begin{align}
    V_{pqrs} & \rightarrow (1/4) \sum_{\ell} \lambda_{\ell}  (u_{pq,\ell}+ u_{qp,\ell} )(u_{sr,\ell} + u_{rs,\ell} ) \\
   & = (1/4) ( V_{pqrs} +   V_{qprs} + V_{pqsr} + V_{qpsr}  )
\end{align}
Using the 8-fold symmetry of \Eq{EightFold}, we have that the Hamiltonian is unchanged under this transform. Note that this approach is essentially a proof that for real-valued orbitals, the skew-Hermitian components can be made to vanish.

For a Hamiltonian spectrally decomposed as described above as
\begin{align}
    H & = H(\tilde{h}) + \sum_{\ell=1}^L \lambda_{\ell}  H(X_{\ell}) H(X_{\ell})
\end{align}
we consider a Trotter decomposition where each term in the product formula implements evolution under one of the terms $H(\tilde{h})$ or $\lambda_{\ell}  H(X_{\ell}) H(X_{\ell})$. The first-order commutator bound on the Trotter error is given by (defining $H_{i=0} := H(\tilde{h})$, $H_{i>0} := \lambda_i H(X_i)H(X_i)$)
\begin{flalign}
    W_1 &= \frac{1}{2}\sum_{i=0}^L \bigg{|}\bigg{|} \sum_{j>i} [H_i, H_j] \bigg{|}\bigg{|}_\eta & \\
    &= \frac{1}{2}\bigg{|}\bigg{|} \sum_{j=1}^L \lambda_j [H(\tilde{h}), H(X_j) H(X_j)] \bigg{|}\bigg{|}_\eta  + \frac{1}{2} \sum_{i=1}^L \bigg{|}\bigg{|} \sum_{j>i}^L \lambda_i \lambda_j [H(X_i) H(X_i), H(X_j) H(X_j)] \bigg{|}\bigg{|}_\eta \nonumber
\end{flalign}
The first term can be expanded using $[A, BC] = [A,B]C + B[A,C]$
\begin{flalign}
    \lambda_j [H(\tilde{h}) , H(X_j) H(X_j)] &= \lambda_j \big{(} [H(\tilde{h}) , H(X_j)] H(X_j) + H(X_j) [H(\tilde{h}) , H(X_j)] \big{)} \nonumber \\
    &= \lambda_j \big{(} H([\tilde{h}, X_j]) H(X_j) + H(X_j) H([\tilde{h}, X_j]) \big{)}.
\end{flalign}
Applying the triangle inequality and H\"older inequality, the bound on the fermionic seminorm of the first term is given by
\begin{flalign}
    \bigg{|}\bigg{|} \sum_{j=1}^L \lambda_j [H(\tilde{h}), H(X_j) H(X_j)] \bigg{|}\bigg{|}_\eta &\leq 2\sum_j |\lambda_j | \cdot \sfs{ [\tilde{h}, X_j] } \cdot \sfs{X_j} . 
\end{flalign}
The second term can be expanded using $[AB, CD] = A[B,C]D + CA[B,D] + [A,C]BD + C[A,D]B$
\begin{flalign}
    \lambda_i \lambda_j [H(X_i) H(X_i), H(X_j) H(X_j)] &= \lambda_i \lambda_j \bigg{(} H(X_i) H([X_i, X_j]) H(X_j) \nonumber \\
    &\quad \quad + H(X_j) H(X_i) H([X_i, X_j]) \\
    &\quad \quad + H([X_i, X_j]) H(X_i) H(X_j) \nonumber \\
    &\quad \quad + H(X_j) H([X_i, X_j]) H(X_i) \bigg{)} \nonumber.
\end{flalign}
Applying the triangle inequality and H\"older inequality, the bound on the fermionic seminorm of the second term is given by
\begin{flalign}
\sum_{i=1}^L \bigg{|}\bigg{|} \sum_{j>i} \lambda_i \lambda_j [H(X_i) H(X_i), H(X_j) H(X_j)] \bigg{|}\bigg{|}_\eta \leq 4 \sum_{i=1, j > i}^L |\lambda_i| \cdot |\lambda_j| \cdot \sfs{[X_i, X_j]} \cdot \sfs{X_i} \cdot \sfs{X_j}.
\end{flalign}
Thus
\begin{flalign}
W_1 \leq \sum_{j=1}^L \bigg{(} |\lambda_j | \cdot \sfs{ [\tilde{h}, X_j] } \cdot \sfs{X_j} \bigg{)} + 2 \sum_{i=1, j > i}^L \bigg{(} |\lambda_i| \cdot |\lambda_j| \cdot \sfs{[X_i, X_j]} \cdot \sfs{X_i} \cdot \sfs{X_j} \bigg{)}.
\end{flalign}
This proves \eq{SpectralBoundMainText} in the main text.

Lastly, we discuss how the above decompositions are related to the spectral decomposition used in \cref{Subsec:SpectralDecomp}. For the relevant special case of the plane wave dual basis (which is a basis of real orbitals) we have $V_{(pq),(sr)}= V_{ps} \delta_{p,q} \delta_{s,r}$.  In other words, $V_{ps}$ is simply the nonzero sub-block of $V_{(pq),(sr)}$.  As such, it is equivalent to diagonalize the smaller matrix $V_{ps}$.

\section{Second-order Trotter error bounds}\label{App:SecondOrderDecomp}

We consider the second-order commutator bounds for a plane wave dual basis Hamiltonian decomposed as $H = H_t + H_v$, with $H_t = H(T)$, and $H_v = H(U) + \sum_l H(X\defindex{l}) H(Y\defindex{l})$. As shown in the main text, the first-order commutator is given by
\begin{equation}
    \begin{aligned}
        [H_t, H_v] = H([T, U]) + \sum_l H([T, X\defindex{l}]) H(Y\defindex{l}) + H(X\defindex{l}) H([T, Y\defindex{l}]).
    \end{aligned}
\end{equation}

The first second-order commutator $[[H_t, H_v], H_t]$ is given by
\begin{equation}
    \begin{aligned}
        [[H_t, H_v], H_t] =& \bigg{[}H([T, U]) + \sum_l H([T, X\defindex{l}]) H(Y\defindex{l}) + H(X\defindex{l}) H([T, Y\defindex{l}])~,~H(T) \bigg{]} \\
        =& H([[T,U],T]) \\
        &\quad + \sum_l \bigg{(} [H([T, X\defindex{l}]) H(Y\defindex{l}), H(T)] + [H(X\defindex{l}) H([T, Y\defindex{l}]), H(T)] \bigg{)} \\
        =& H([[T,U],T]) + \sum_l \bigg{(} H([T, X\defindex{l}]) H([Y\defindex{l}, T]) +  H([[T, X\defindex{l}], T]) H(Y\defindex{l}) \\
        & \quad + H(X\defindex{l}) H([[T, Y\defindex{l}], T]) + H([X\defindex{l}, T]) H([T, Y\defindex{l}]) \bigg{)}.
    \end{aligned}
\end{equation}
The fermionic seminorm of this expression is bounded by
\begin{equation}\label{Eq:TVT_bound}
    \begin{aligned}
        \big{|}\big{|} [[H_t, H_v], H_t] \big{|}\big{|}_\eta &\leq  \sfs{[[T,U],T]} + \sum_l \bigg{(} 2\sfs{[T, X\defindex{l}] } \cdot \sfs{[Y\defindex{l}, T]} +  \sfs{[[T, X\defindex{l}], T]} \cdot \sfs{Y\defindex{l}} \\
        &\quad \quad + \sfs{X\defindex{l}} \cdot \sfs{[[T, Y\defindex{l}], T]} \bigg{)}.
    \end{aligned}
\end{equation}
The other second-order commutator $[[H_t, H_v], H_v]$ is given by
\begin{equation}
    \begin{aligned}
        [[H_t, H_v], H_v] &= \bigg{[} H([T, U]) + \sum_l H([T, X\defindex{l}]) H(Y\defindex{l}) + H(X\defindex{l}) H([T, Y\defindex{l}])~~, \\
        &\quad \quad \quad \quad H(U) + \sum_m H(X\defindex{m}) H(Y\defindex{m})     \bigg{]} \\
        &= H([[T,U],U]) + t_1 + t_2 + t_3 + t_4 + t_5 
      \end{aligned}
\end{equation}      
where
\begin{align}
    t_1  &= \sum_l \big{[} H([T, X\defindex{l}]) H(Y\defindex{l}), H(U)  \big{]} \\
    t_2 &= \sum_l \big{[}  H(X\defindex{l}) H([T, Y\defindex{l}]), H(U)  \big{]} \\
    t_3    &= \sum_m \big{[}  H([T, U]), H(X\defindex{m}) H(Y\defindex{m})  \big{]} \\
    t_4    &= \sum_{l,m} \big{[} H([T, X\defindex{l}]) H(Y\defindex{l}),  H(X\defindex{m}) H(Y\defindex{m})    \big{]}  \\
    t_5    &=\sum_{l,m} \big{[}  H(X\defindex{l}) H([T, Y\defindex{l}]) ,  H(X\defindex{m}) H(Y\defindex{m})    \big{]}.
\end{align}
We evaluate these terms separately:
    \begin{align}
t_1 =& \sum_{l} H([T, X\defindex{l}]) H([Y\defindex{l}, U]) + H([[T, X\defindex{l}], U]) H(Y\defindex{l}) \\
t_2 = &  \sum_{l} H(X\defindex{l}) H([[T, Y\defindex{l}], U]) + H([X\defindex{l}, U]) H([T, Y\defindex{l}]) \\
t_3 =& \sum_m H([[T, U], X\defindex{m}]) H(Y\defindex{m}) + H(X\defindex{m}) H([[T, U], Y\defindex{m}]) \\
t_4 =& \sum_{l,m} H([T, X\defindex{l}]) H([Y\defindex{l}, X\defindex{m}]) H(Y\defindex{m}) + H([[T, X\defindex{l}], X\defindex{m}]) H(Y\defindex{l}) H(Y\defindex{m}) \\
        & + H(X\defindex{m}) H([T, X\defindex{l}]) H([Y\defindex{l}, Y\defindex{m}]) + H(X\defindex{m}) H([[T, X\defindex{l}], Y\defindex{m}] H(Y\defindex{l}) \nonumber \\
t_5 =& \sum_{l,m} H(X\defindex{l}) H([[T, Y\defindex{l}], X\defindex{m}]) H(Y\defindex{m}) + H([X\defindex{l}, X\defindex{m}]) H([T, Y\defindex{l}]) H(Y\defindex{m}) \\
        & + H(X\defindex{m}) H(X\defindex{l}) H([[T, Y\defindex{l}], Y\defindex{m}]) + H(X\defindex{m}) H([X\defindex{l}, Y\defindex{m}]) H([T, Y\defindex{l}]) \nonumber
    \end{align}
Note that the $4\mathrm{th}$ term of $t_4$ can be combined with the $1\mathrm{st}$ term of $t_5$ was follows
\begin{align}  \label{eq:ExtraSimplify}
   & \sum_{l,m}H(X\defindex{m}) H([[T, X\defindex{l}], Y\defindex{m}] H(Y\defindex{l}) + H(X\defindex{l}) H([[T, Y\defindex{l}], X\defindex{m}]) H(Y\defindex{m}) \nonumber \\
   &= \sum_{l,m} H(X\defindex{m}) H([[T, X\defindex{l}], Y\defindex{m}] H(Y\defindex{l}) + H(X\defindex{m}) H([[T, Y\defindex{m}], X\defindex{l}]) H(Y\defindex{l}) \nonumber \\
   &= \sum_{l,m} H(X\defindex{m}) \left\{ H( [[T, X\defindex{l}], Y\defindex{m}] + [[T, Y\defindex{m}], X\defindex{l}]) \right\} H(Y\defindex{l})  
\end{align}

We can then bound the fermionic seminorm of the commutator (ignoring \cref{eq:ExtraSimplify}) by

\begin{flalign}\label{Eq:TVV_bound}
    \big{|}\big{|} [[H_t, H_v]&, H_v] \big{|}\big{|}_\eta \leq \sfs{[[T, U], U]} \nonumber \\
        &+ \sum_l \bigg{(}  \sfs{[T, X\defindex{l}]} \cdot \sfs{[Y\defindex{l}, U]} + \sfs{[[T, X\defindex{l}], U]} \cdot \sfs{Y\defindex{l}} \nonumber \\
         &+ \sfs{X\defindex{l}} \cdot \sfs{[[T, Y\defindex{l}], U]} + \sfs{[X\defindex{l}, U]} \cdot \sfs{[T, Y\defindex{l}]} \nonumber \\
         &+ \sfs{ [[T, U], X\defindex{l}] } \cdot \sfs{Y\defindex{l}} + \sfs{ [[T, U], Y\defindex{l}] } \cdot \sfs{X\defindex{l}} \bigg{)} \nonumber \\
        &+ \sum_{l,m} \bigg{(} \sfs{[T, X\defindex{l}]} \cdot \sfs{[Y\defindex{l}, X\defindex{m}]} \cdot \sfs{Y\defindex{m}} + \sfs{[[T, X\defindex{l}], X\defindex{m}]} \cdot \sfs{Y\defindex{l}} \cdot \sfs{Y\defindex{m}} \nonumber \\
        &+  \sfs{X\defindex{m}} \cdot \sfs{[T, X\defindex{l}]} \cdot \sfs{[Y\defindex{l}, Y\defindex{m}]} + \sfs{X\defindex{m}} \cdot \sfs{[[T, X\defindex{l}], Y\defindex{m}]} \cdot \sfs{Y\defindex{l}} \nonumber \\
        &+ \sfs{X\defindex{l}} \cdot \sfs{[[T, Y\defindex{l}], X\defindex{m}]} \cdot \sfs{Y\defindex{m}} + \sfs{[X\defindex{l}, X\defindex{m}]} \cdot \sfs{[T, Y\defindex{l}]} \cdot \sfs{Y\defindex{m}} \nonumber \\
        &+ \sfs{X\defindex{m}} \cdot \sfs{X\defindex{l}} \cdot \sfs{[[T, Y\defindex{l}], Y\defindex{m}]} + \sfs{X\defindex{m}} \cdot \sfs{[X\defindex{l}, Y\defindex{m}]} \cdot \sfs{[T, Y\defindex{l}} \bigg{)} 
    \end{flalign}

And using the simplification of \cref{eq:ExtraSimplify} we get
\begin{flalign}\label{Eq:TVV_bound2}
    \big{|}\big{|} [[H_t, H_v]&, H_v] \big{|}\big{|}_\eta \leq \sfs{[[T, U], U]} \nonumber \\
        &+ \sum_l \bigg{(}  \sfs{[T, X\defindex{l}]} \cdot \sfs{[Y\defindex{l}, U]} + \sfs{[[T, X\defindex{l}], U]} \cdot \sfs{Y\defindex{l}} \nonumber \\
         &+ \sfs{X\defindex{l}} \cdot \sfs{[[T, Y\defindex{l}], U]} + \sfs{[X\defindex{l}, U]} \cdot \sfs{[T, Y\defindex{l}]} \nonumber \\
         &+ \sfs{ [[T, U], X\defindex{l}] } \cdot \sfs{Y\defindex{l}} + \sfs{ [[T, U], Y\defindex{l}] } \cdot \sfs{X\defindex{l}} \bigg{)} \nonumber \\
        &+ \sum_{l,m} \bigg{(} \sfs{[T, X\defindex{l}]} \cdot \sfs{[Y\defindex{l}, X\defindex{m}]} \cdot \sfs{Y\defindex{m}} + \sfs{[[T, X\defindex{l}], X\defindex{m}]} \cdot \sfs{Y\defindex{l}} \cdot \sfs{Y\defindex{m}} \nonumber \\
        &+  \sfs{X\defindex{m}} \cdot \sfs{[T, X\defindex{l}]} \cdot \sfs{[Y\defindex{l}, Y\defindex{m}]} \nonumber + \sfs{[X\defindex{l}, X\defindex{m}]} \cdot \sfs{[T, Y\defindex{l}]} \cdot \sfs{Y\defindex{m}} \nonumber \\ \nonumber
        &+ \sfs{X\defindex{m}} \cdot \sfs{X\defindex{l}} \cdot \sfs{[[T, Y\defindex{l}], Y\defindex{m}]} + \sfs{X\defindex{m}} \cdot \sfs{[X\defindex{l}, Y\defindex{m}]} \cdot \sfs{[T, Y\defindex{l}}  \\
     &  + \sfs{X\defindex{l}} \left\{ \sfs{[[T, Y\defindex{l}], X\defindex{m}] + [[T, X\defindex{m}], Y\defindex{l}] } \right\} \sfs{Y\defindex{m}} \bigg{)}.
    \end{flalign}

\section{Second-order commutator bounds for plane wave dual decompositions}\label{App:SpecificSecondOrderBounds}

In this Appendix, we apply the formulae for the second-order commutator bound for a general decomposition to the decompositions introduced in the main text. \\

Comparing the spectral decomposition in the main text to \Eq{TVT_bound}, \Eq{TVV_bound}, we observe that
$X\defindex{i}:= \lambda_i v\defindex{i}$, $Y\defindex{i} := v\defindex{i}$ and note that $[X\defindex{i}, X\defindex{j}] = [Y\defindex{i}, Y\defindex{j}] = [X\defindex{i}, Y\defindex{j}] = 0~\forall~i,j$. The second-order commutator bounds are then given by
\begin{equation}
\begin{aligned}
    \big{|} \big{|} [[H_t, H_v], H_t] \big{|} \big{|}_\eta &\leq \sfs{[[T, U], T]} + 2 \sum_{i}  |\lambda_i| \bigg{(} \sfs{[[T, v\defindex{i}], T]} \cdot \sfs{v\defindex{i}} + \sfs{[T, v\defindex{i}]}^2 \bigg{)}
\end{aligned}
\end{equation}
\begin{flalign}
    &\big{|} \big{|} [[H_t, H_v], H_v] \big{|} \big{|}_\eta \\ 
    &\leq \sfs{[[T, U], U]} \nonumber \\
    &+ 2 \sum_{i} |\lambda_i| \bigg{(} \sfs{[T, v\defindex{i}]} \cdot \sfs{[v\defindex{i}, U]} + \sfs{[[T, v\defindex{i}], U]} \cdot \sfs{v\defindex{i}} + \sfs{ [[T, U], v\defindex{i}] } \cdot \sfs{v\defindex{i} } \bigg{)} \nonumber \\
    &\quad \quad + 4 \sum_{i,j} |\lambda_i| |\lambda_j| \bigg{(} \sfs{[[T, v\defindex{i}], v\defindex{j}]} \cdot \sfs{v\defindex{i}} \cdot \sfs{v\defindex{j}} \bigg{)}. \nonumber
\end{flalign}\\

Comparing the Cholesky decomposition in the main text to \Eq{TVT_bound}, \Eq{TVV_bound}, we observe that
$X\defindex{i}:= L\defindex{i}$, $Y\defindex{i} := L\defindex{i}$ and note that $[X\defindex{i}, X\defindex{j}] = [Y\defindex{i}, Y\defindex{j}] = [X\defindex{i}, Y\defindex{j}] = 0~\forall~i,j$. The second-order commutator bounds are then given by
\begin{equation}
\begin{aligned}
    \big{|} \big{|} [[H_t, H_v], H_t] \big{|} \big{|}_\eta &\leq \sfs{[[T, U], T]} + 2 \sum_{i}  \bigg{(} \sfs{[[T, L\defindex{i}], T]} \cdot \sfs{L\defindex{i}} + \sfs{[T, L\defindex{i}]}^2 \bigg{)}
\end{aligned}
\end{equation}
\begin{flalign}
    &\big{|} \big{|} [[H_t, H_v], H_v] \big{|} \big{|}_\eta \\ 
    &\leq \sfs{[[T, U], U]} \nonumber \\
    &+ 2 \sum_{i}  \bigg{(} \sfs{[T, L\defindex{i}]} \cdot \sfs{[L\defindex{i}, U]} + \sfs{[[T, L\defindex{i}], U]} \cdot \sfs{L\defindex{i}} + \sfs{ [[T, U], L\defindex{i}] } \cdot \sfs{L\defindex{i} } \bigg{)} \nonumber \\
    &\quad \quad + 4 \sum_{i,j}  \bigg{(} \sfs{[[T, L\defindex{i}], L\defindex{j}]} \cdot \sfs{L\defindex{i}} \cdot \sfs{L\defindex{j}} \bigg{)}. \nonumber
\end{flalign}\\

Comparing the cosine decomposition in the main text to  \Eq{TVT_bound}, \Eq{TVV_bound}, we observe that $H(X\defindex{q}) H(Y\defindex{q}) := H(A\defindex{\nu}) H(A\defindex{\nu})$ such that $ A\defindex{\nu} \in \{C\defindex{\nu}, S\defindex{\nu} \}$. Because these are diagonal matrices, all of the commutators between different $X$ matrices, different $Y$ matrices, and between $X$ and $Y$ matrices vanish. The resulting commutator bounds are given by
\begin{equation}
\begin{aligned}
    \big{|}\big{|} [[H_t, H_v], H_t] \big{|}\big{|}_\eta &\leq \sfs{[[T, U],T]} + 2 \sum_{\nu \neq 0} \sum_{A \in \{C,S \}} \bigg{(} \sfs{[[T, A\defindex{\nu}], T]} \cdot \sfs{A\defindex{\nu}} + \sfs{[T, A\defindex{\nu}]}^2 \bigg{)}
\end{aligned}
\end{equation}
\begin{equation}
\begin{aligned}
    \big{|}\big{|} [[H_t, H_v], H_v] \big{|}\big{|}_\eta &\leq \sfs{[[T, U], U]} \\
    &+ 2\sum_{\nu \neq 0} \sum_{A \in \{ C, S \}} \bigg{(} \sfs{[T, A\defindex{\nu}]} \cdot\sfs{[U, A\defindex{\nu}]} \\
    &+ \sfs{[[T, A\defindex{\nu}], U]} \cdot \sfs{A\defindex{\nu}} + \sfs{ [[T, U], A\defindex{\nu}] } \cdot \sfs{ A\defindex{\nu} }  \bigg{)} \\
    &+ 4 \sum_{\nu, \xi \neq 0} \sum_{A \in \{C\defindex{\nu},S\defindex{\nu}\}} \sum_{B \in \{C\defindex{\xi},S\defindex{\xi}\}} \bigg{(} \sfs{[[T, A], B]} \cdot \sfs{A} \cdot \sfs{B} \bigg{)}.
\end{aligned}
\end{equation}\\

\section{Bound computation details}\label{App:ComputationDetails}

In this section we outline how the commutator bounds discussed in the main text were calculated. For the case of the first-order Fermionic commutator bounds, we must evaluate $\sum_{i,j} \sum_{p,q} [T_{ij} a_i^\dag a_j, V_{pq} n_p n_q]$. Commutators where $p,q$ are both distinct from $i,j$ trivially commute, leading to $\mathcal{O}(N^3)$ distinct commutators to store. As a result, the time cost for the algorithm is $\mathcal{O}(N^4)$ and the memory cost is $\mathcal{O}(N^3)$. For the second-order bounds, there are $\mathcal{O}(N^4)$ terms to store, and the algorithm has time cost $\mathcal{O}(N^6)$. We store the resulting commutators, collect like terms, and then apply the triangle inequality to the sum.

Pauli commutators can be evaluated in a similar manner. After evaluating the fermionic commutators to calculate the error operator, we apply the Jordan-Wigner transform to obtain the error operator written as a sum of tensor products of Pauli operators. Local fermionic operators are mapped to $\mathcal{O}(N)$-local Jordan-Wigner operators, which increases the memory required by a factor of $\mathcal{O}(N)$.

The cosine, Cholesky, and spectral bounds can be calculated by storing in memory the $\mathcal{O}(N)$ diagonal $N \times N$ coefficient matrices, leading to a memory cost of $\mathcal{O}(N^2)$ if only storing the diagonal elements. The calculation of $\big{|}\big{|} [[H_t, H_v], H_v] \big{|}\big{|}_\eta$ requires $\mathcal{O}(N^2)$ loop passes, where the dominant costs in each pass are multiplication and diagonalization of the coefficient matrices. These operations have a cost of approximately $\mathcal{O}(N^3)$. As a result, the time cost of the algorithm is approximately $\mathcal{O}(N^5)$.

\begin{table*}[]
    \centering
    \begin{tabular}{c|c|c}
         Approach & Analytic runtime & Empirical runtime \\
          & $\mathcal{O}(N^\alpha)$ & $\mathcal{O}(N^\alpha)$ \\
         \hline
         Fermionic commutator~\cite{Kivlichan2020improvedfault} & 6 & $3.61$ \\
         SHC bound \cite{su2020nearly}  & 3 & $2.95$  \\
        Spectral decomp. [This work] & 5 & $3.76$ \\
        Cholesky decomp. [This work] & 5 & $3.79$  \\
         Cosine decomp. [This work] & 5 & $3.35$
    \end{tabular}
\caption{A comparison between the analytic and empirical runtime scalings of the methods used in this work to calculate second-order Trotter bounds. Empirical scalings were determined from numerical simulations performed on a system with 49 electrons in 98 -- 512 spin-orbitals (98 -- 200 spin-orbitals for the fermionic bound). The time taken, $T$, was fitted to the function $T = kN^\alpha$ to determine the empirical runtime scalings. We attribute the impressive empirical performance of the fermionic commutator bound scaling to a highly optimised implementation present in OpenFermion~\cite{mcclean2020openfermion}. However, we note that this approach is still limited by the large memory requirement of the fermionic commutator bound.}
    \label{tab:Scalings}
\end{table*}

\section{Projected Pauli bounds}\label{App:ProjectedPauli}
\begin{figure}[!h]
\begin{center}
\includegraphics[width=0.9\columnwidth]{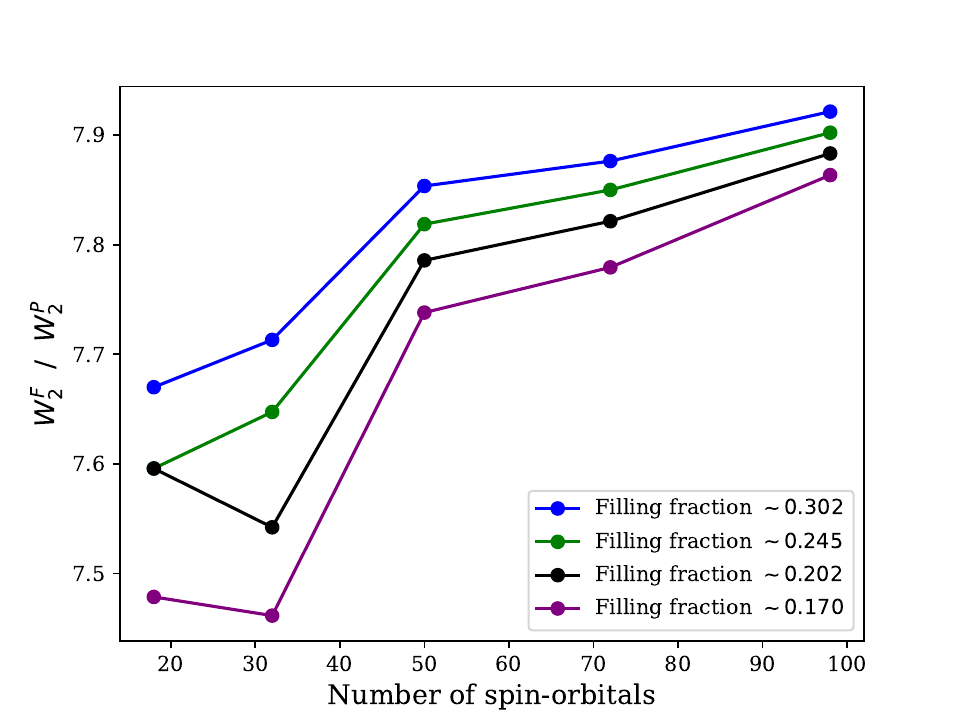}
\caption{The ratio between the second-order fermionic commutator bound $W_2^F$ and the second-order Pauli commutator bound $W_2^P$, as a function of the number of spin-orbitals used, for a homogeneous electron gas system with $r_s=5$. The number of electrons in each calculation is varied, in order to match the specified filling fraction as closely as possible.}
\label{Fig:ProjectedPauli}
\end{center}
\end{figure}

The high memory requirements of the Pauli commutator bound make it impractical to calculate $W_2^{\mathrm{Pauli}}$ for $N \geq 128$. However, it is evident from Fig.~\ref{Fig:fixed_electrons} that the Pauli bounds appear to be only a constant factor better than the fermionic commutator bounds. As a result, we can estimate the Pauli commutator bounds for larger $N$ values, using the available fermionic commutator datapoints. We wish to predict $W_2^{\mathrm{Pauli}}$ at $N=162, 200, 242, 288$, which for 49 electrons, corresponds to a filling fraction of $0.302, 0.245, 0.202, 0.170$, respectively. In Fig.~\ref{Fig:ProjectedPauli}, we plot the ratio between the second-order fermionic commutator bound $W_2^{\mathrm{Ferm}}$, and $W_2^{\mathrm{Pauli}}$, for a range of $N$ values, varying the number of electrons such that the filling fraction is kept approximately constant. We observe that as the number of orbitals used increases, the ratio gradually increases. The ratio decreases as the filling fraction decreases. As a result, we assume that at $N=162, 200, 242, 288$, the Pauli bound outperforms the fermionic bound by roughly a factor of 8.

\section{Phase estimation resource costs}\label{App:PhaseEstErrorBudget}

We first discuss the resources used to implement a single Trotter step of time evolution. When the number of Trotter steps is large, the difference in gate count per Trotter step between implementing $e^{i\frac{t}{2}H_v}e^{itH_t}e^{i\frac{t}{2}H_v}$ and $e^{i\frac{t}{2}H_t}e^{itH_v}e^{i\frac{t}{2}H_t}$ is negligible. The final term of each Trotter step can be merged with the first term of the next, so that each Trotter step contains one implementation of $e^{iH_t}$ and one of $e^{iH_v}$. The difference in total gate count between these two approaches is thus determined by the difference in Trotter error of the orderings.

For the uniform electron gas, $H_v = \sum_{p \neq q} V_{pq} n_p n_q$ contains $N(N-1)/2$ terms, and so can be implemented by an equivalent number of arbitrary angle $Z$ rotations. However, the translational invariance of Jellium leads many of these rotations to be of the same angle. As discussed in Ref.~\cite{Kivlichan2020improvedfault}, these rotations can be implemented via HWP in groups of size $N/2$. In practice, we counted the multiplicity of the terms in $V_{pq}$, and used HWP to reduce the number of arbitrary rotations required. This contributes an $\mathcal{O}(N^2 \log(\epsilon^{-1}))$ gate complexity to each Trotter step.  Low and Wiebe~\cite{low2018hamiltonian} proposed an alternative approach that would need only $\mathcal{O}(N \log( N ) \log(\epsilon^{-1}))$ gates, but with a significant constant factor overhead that makes it more expensive in the regime considered here.

Changing from the plane wave dual to the plane wave basis can be accomplished using either the fermionic fast Fourier transform (FFFT, when the lattice sides are a power of two)~\cite{verstraete2009quantumcircuits, ferris2014fouriertransform, babbush2018planewaves, Kivlichan2020improvedfault}, or using Givens rotation circuits~\cite{wecker2015hubbard, kivlichan2018linear}. These approaches have similar costs for Jellium~\cite{Kivlichan2020improvedfault}. The FFFT has a recursive structure, and requires $\frac{L}{2} \mathrm{log}_2(L)$ non-Clifford gates when applied to $L$ qubits. The FFFT must be applied multiple times when changing the basis of a grid in multiple dimensions. For a $L_x \times L_y$ spinful lattice, we require $2L_x$ applications of the FFFT on $L_y$ qubits, and $2L_y$ applications of the FFFT on $L_x$ qubits (for a $d$-dimensional spinful lattice of side $L$, we require $2dL^{d-1}$ applications of the FFFT)~\cite{Kivlichan2020improvedfault}. Ref.~\cite{Kivlichan2020improvedfault} determined that implementing the FFFT requires 26 $T$ gates for 8 qubits, and 81 $T$ gates for 16 qubits. Givens rotations can be used to perform a single-particle orbital basis change, regardless of whether the number of orbitals considered is a power of 2. We follow the approach outlined in Ref.~\cite{Kivlichan2020improvedfault}. A single Givens rotation requires two non-Clifford gates, in the form of two arbitrary rotations (by the same angle). A basis change on $M$ qubits requires $M \choose 2$ Givens rotations. As with the FFFT, we perform the Givens rotations a number of times to change basis in multiple dimensions. For an $L_x \times L_y$ spinful lattice, we require $2L_x$ implementations of the basis change on $L_y$ qubits, and $2L_y$ implementations of the basis change on $L_x$ qubits. For the former case (with corresponding changes for the latter), we require $2L_x \times 2 \times$~$L_y \choose 2$ arbitrary rotations. These can be parallelised into $L_y \choose 2$ groups of size $4L_x$. The T/Toffoli cost of implementing these arbitrary rotations can be reduced using Hamming weight phasing.

Rotating into the plane wave basis diagonalises the kinetic operator, enabling us to implement it with $N$ arbitrary rotations in the worst case. As it is efficient to classically diagonalise the kinetic coefficient matrix $T_{pq}$, we can determine the multiplicity of each eigenvalue, and then use HWP to reduce the number of arbitrary rotations required. Overall, implementing this contributes a cost $\mathcal{O}(N \log( N) \log(\epsilon^{-1}) )$ per Trotter step.

To perform phase estimation we must implement not just a circuit approximating $e^{iHt}$, but a circuit that approximates $e^{iHt}$ controlled on the state of an ancillary register. We can implement a controlled arbitrary rotation at double the cost of the un-controlled operation~\cite{nielsen2002quantum}. However, Ref.~\cite{wecker2015hubbard} introduced an approach known as directionally controlled phase estimation, that reduces the cost of controlled time evolution to be the same as the uncontrolled circuit, when implemented with symmetric product formulae (this approach was elaborated upon further in Refs.~\cite{reiher2017elucidating,Kivlichan2020improvedfault}). The key insight is that one instance of $U_2(t)$ can be used to implement $[\ket{0}_a \bra{0}_a \otimes U_2(-t) + \ket{1}_a \bra{1}_a \otimes U_2(t)]$, which for the purposes of phase estimation is equivalent to performing $[\ket{0}_a \bra{0}_a \otimes I + \ket{1}_a \bra{1}_a \otimes U_2(2t)]$. In addition to halving the number of arbitrary rotations required, this optimization effectively doubles the time duration used for phase estimation. We use an adaptive variant of phase estimation that uses a single ancilla qubit~\cite{berry2009phaseestimation}. As discussed in Ref.~\cite{Kivlichan2020improvedfault}, this approach uses $N_{PE}$ applications of directionally controlled phase estimation to learn the energy eigenvalue to a root mean squared error of 
\begin{equation}
    \Delta_{PE} \approx \frac{0.76 \pi}{N_{PE} t}.
\end{equation}
We note that this formula includes the reduction from $t \rightarrow 2t$ due to the use of directionally controlled phase estimation.  The Trotter error contributes an error $\Delta_{TS}= W t^2$ where $W$ is the commutator bound constant.  A third source of error of error are synthesis errors $\Delta_{\mathrm{syn}}=\mathcal{O}(\log(N_R/\epsilon))$ where $N_R$ is the number of arbitrary $Z$ axis rotations in the algorithm. We distribute errors between these three sources using the approach outlined in Appendix F of Ref.~\cite{campbell2020early}.

\begin{table}
\begin{tabular}{cccc|c|cc|c}
  &  & Filling &  & Error  &   &  & Aggregated  \\ 
  &  & fraction & Size & constant  & Tof gates  & T gates & T count  \\ 
   $r_s$ & $\eta$  & $\eta / 2 L_X L_ Y  $ & $L_X \times L_Y$ & $W_2$  & $N_{\mathrm{tof}}$ & $N_{T}$ & $N_T+4 N_{\mathrm{tof}}$  \\ \hline
   5 & 49 & 0.10 & 16 $\times$ 16 & $2.89 \times 10^4$   & $9.7 \times 10^9$ & $1.8 \times 10^{11}$ & $2.2 \times 10^{11}$  \\
   5 & 49 & 0.17 & 12$\times$12 & $5.18 \times 10^3$   & $1.5 \times 10^9$ & $2.6\times 10^{10}$ & $3.2 \times 10^{10}$ \\
   5 & 49 & 0.19 & 16$\times$8 & $3.44 \times 10^3$    & $8.4 \times 10^8$ & $1.4 \times 10^{10}$ & $1.7 \times 10^{10}$  \\
   5 & 49 & 0.38 & 8$\times$8 & 356   & $6.8 \times 10^7$  & $1.1 \times 10^9$ & $1.3 \times 10^9$  \\  \hline
   10 & 10 & 0.02 & 16$\times$16 & 604   & $1.6 \times 10^{10}$  & $2.8 \times 10^{11}$ & $3.4 \times 10^{11}$  \\
   10 & 10 & 0.03 & 12$\times$12 & 290   & $3.9 \times 10^9$ & $7.0 \times 10^{10}$ & $8.6 \times 10^{10}$  \\
   10 & 10 & 0.04 & 16$\times$8 & 262   & $2.5 \times 10^9$  & $4.4 \times 10^{10}$ & $5.4 \times 10^{10}$  \\
   10 & 10 & 0.08 & 8$\times$8 & 103   & $3.9 \times 10^8$  & $6.6 \times 10^9$ & $8.1 \times 10^9$  \\
   10 & 49 & 0.10 & 16$\times$16 & $7.20 \times 10^3$    & $4.8 \times 10^9$ & $8.6 \times 10^{10}$ & $1.1 \times 10^{11}$  \\
   10 & 49 & 0.17 & 12$\times$12 & $1.29 \times 10^3$    & $7.6 \times 10^8$ & $1.3 \times 10^{10}$ & $1.6 \times 10^{10}$  \\
   10 & 49 & 0.19 & 16$\times$8 & 857    & $4.2 \times 10^8$ & $6.9 \times 10^9$  & $8.6 \times 10^9$  \\
   10 & 49 & 0.38 & 8$\times$8 & 89    & $3.4 \times 10^7$ & $5.1 \times 10^8$ & $6.5 \times 10^8$ \\ \hline
\end{tabular} \caption{Resource estimates for phase estimation of Jellium. We consider an energy error budget of $\delta=1$ mHa per electron. The `Error constant' is obtained using the most performant of the bounds introduced in this work, for each system (this is either the Cholesky or cosine decomposition for all datapoints shown). We use 16 additional qubits (14 for Hamming weight phasing, one for phase estimation, and one for gate synthesis). Four $T$ gates can be used to implement a Toffoli gate, so the total aggregated $T$ count for the algorithm is  $N_{T}+4N_{\mathrm{tof}}$.}  \label{tab:JelliumPhaseEstData288}
\end{table}

\section{Comparison to Qubitization}\label{App:Qubitization}
The approaches presented in this work for performing Trotter-based phase estimation of systems in a plane wave dual basis can be compared to the approach introduced in Ref.~\cite{babbush2018EncodingElectronicSpectra}, which considered a qubitization-based approach to phase estimation. This approach divides the Hamiltonian into a linear combination of unitary operators $H = \sum_a h_a H_a$ (with $H_a$ unitary, e.g. Pauli strings), and uses circuits to `block encode' $H$ in a subspace of a Hilbert space enlarged by additional ancilla qubits~\cite{Low2019hamiltonian}. By repeating the block encoding procedure, one can perform a quantum walk, the eigenvalues of which are related to the eigenvalues of the Hamiltonian, without approximation errors~\cite{berry2018improved, poulin2018spectral}. One can then perform phase estimation directly on this walk operation~\cite{berry2018improved, poulin2018spectral}. The $T$ cost of this qubitization approach for Jellium is given by Eq.(54) in Ref.~\cite{babbush2018EncodingElectronicSpectra} as
\begin{equation}
    \frac{24 \sqrt{2} \pi \lambda N}{\delta}
\end{equation}
where for a Hamiltonian written as $H = \sum_a h_a H_a$ (with $|| H_a || = 1$), $\lambda = \sum_a |h_a|$, $N$ is the number of spin-orbitals, and $\delta$ is the target energy error. The number of logical ancilla qubits required is given by Eq.(55) of Ref.~\cite{babbush2018EncodingElectronicSpectra}
\begin{equation}
    \mathrm{log}_2\bigg{(}\frac{4 \sqrt{2} \pi \lambda^3 N^5}{\delta^3} \bigg{)}.
\end{equation}
It is interesting to consider how the gate count scales as a function of $r_s$. We have that for Jellium, $\lambda = \lambda_t + \lambda_v$. For a dim-$d$ system, we can see directly from the Hamiltonian coefficients in Eq.(\ref{Eq:Coefficients}) (using that $\Omega \propto \eta r_s^d$) that
\begin{align}
    \lambda_t &\sim \frac{1}{\eta^{2/d} r_s^2} \\
    \lambda_v &\sim \eta^{\frac{2}{d} - 1} r_s^{2- d}
\end{align}
In these expressions we have implicitly assumed that $N$ is held constant. As a result, 
in 2D $\lambda$ scales as $\mathcal{O}(\eta^{-1} r_s^{-2}) + \mathcal{O}(1)$. This can be contrasted with our second-order Trotter approach. We have that
\begin{align}
    W_2 &\sim || [[H_t, H_v], H_t]|| + || [[H_t, H_v], H_v]|| \nonumber \\
    &< \lambda_t^2 \lambda_v + \lambda_t \lambda_v^2 \nonumber \\
    &\sim \mathcal{O}(\eta^{-(2/d + 1)} r_s^{-(2+d)}) + \mathcal{O}(\eta^{(2/d -2)} r_s^{2-2d})
\end{align}
For $d=2$, $W_2 = \mathcal{O}(\eta^{-2} r_s^{-4}) + \mathcal{O}(\eta^{-1} r_s^{-2})$. We note that this bound on the Trotter error may be very loose (in terms of the scaling with the number of electrons), as it does not use commutativity of terms in the Hamiltonian or the fermionic seminorm (c.f. Eq.~\ref{Eq:SHC_asymptotic}).

If we fix $\eta, N$ and vary $r_s$, we see that for $d=2$ and large $r_s$, the cost of qubitization is independent of $r_s$, while our Trotter-based approach scales as $ \mathcal{O}(r_s^{-1})$. Thus, the cost of Trotter-based approaches in 2D reduce as the value of $r_s$ is increased, while the cost of qubitization is roughly independent of $r_s$. This is evident in the results presented in Table~\ref{tab:JelliumPhaseEstDataComparison}.

\end{document}